\documentclass{aa}
\usepackage{graphicx}
\usepackage{txfonts}
\usepackage{natbib}

\bibpunct[]{(}{)}{;}{a}{}{,}
\def\ltsima{$\; \buildrel < \over \sim \;$}
\def\simlt{\lower.5ex\hbox{\ltsima}}
\def\gtsima{$\; \buildrel > \over \sim \;$}
\def\simgt{\lower.5ex\hbox{\gtsima}}

\def\cm2{\mbox{$\mbox{cm}^{-2}$}}
\def\cm3{\mbox{$\mbox{cm}^{-3}$}}
\def\h2{\mbox{$_{\mbox{\tiny H2}}$}}

\begin{document}
   \title{The initial conditions of star formation\\ 
in the Ophiuchus main cloud:\\
Kinematics of the protocluster condensations\thanks{Based on observations carried out with the 
IRAM 30-meter telescope. IRAM is supported by INSU/CNRS (France), MPG (Germany), and IGN (Spain).}}

   \author{Ph. Andr\'e\inst{1,2}
          \and
          A. Belloche\inst{1,3}
          \and
	  F. Motte\inst{1,2}
          \and
          N. Peretto\inst{1,4}
          }

   \offprints{pandre@cea.fr}

   \institute{CEA/DSM/DAPNIA, Service d'Astrophysique, C.E. Saclay,
              Orme des Merisiers, 91191 Gif-sur-Yvette, France
         \and
		Laboratoire AIM, Unit\'e Mixte de Recherche CEA/DSM -- CNRS 
		  -- Universit\'e Paris Diderot, C.E. Saclay, France
	 \and
              Max-Planck-Institut f\"ur Radioastronomie, Auf dem H\"ugel 69,
              53121 Bonn, Germany
 	 \and
	     Physics  \& Astronomy Department, University of Manchester, P.O. Box 88, 
	     Manchester M60 1QD, UK
             }

   \date{Received  6 March 2007 /  Accepted 7 June 2007 }

   \abstract
  {The earliest phases of clustered star formation and the origin of the stellar initial mass function (IMF)
  are currently much debated. In one school of thought the IMF of embedded clusters 
  is entirely determined by turbulent fragmentation at the prestellar stage of star formation, while in a major 
  alternative view it results from dynamical interactions and competitive accretion at the protostellar stage.} 
  {In an effort to discriminate between these two pictures for the origin of the IMF, we investigated 
  the internal and relative motions of starless condensations and protostars
  previously detected by us in the dust continuum at 1.2~mm 
  in the L1688 protocluster of the Ophiuchus molecular cloud complex.
  The starless condensations have a mass spectrum 
  resembling the IMF and are therefore likely representative of the initial stages of star formation in the protocluster.}
  {We carried out detailed molecular line observations, including some N$_2$H$^+$(1-0) mapping, of the 
  Ophiuchus protocluster condensations using the IRAM 30m telescope. 
  }
  {We measured subsonic or at most transonic levels of internal turbulence within the condensations, implying virial masses 
  which generally agree within a factor of $\sim 2$ with the masses derived from the 1.2~mm dust continuum.
  This supports the notion that most 
  of the L1688 starless condensations are gravitationally bound and prestellar in nature. 
  We detected the classical spectroscopic signature of infall motions in CS(2--1), CS(3--2), H$_2$CO($2_{12} - 1_{11}$), 
  and/or HCO$^+$(3--2) toward six condensations, and obtained tentative infall signatures toward 10 other condensations.
  In addition, we measured a global one-dimensional velocity dispersion of less than $0.4$ km~s$^{-1}$ (or twice 
  the sound speed) between condensations. The small relative velocity dispersion implies that, in general, the 
  condensations do not have time to interact with one another before evolving into pre-main sequence 
  objects.}
  {Our observations support the view that the IMF is partly determined by cloud fragmentation at the 
  prestellar stage. Competitive accretion is unlikely to be the dominant mechanism at the protostellar stage
  in the Ophiuchus protocluster, but it  may possibly 
  govern the growth of starless, self-gravitating 
  condensations initially produced by gravoturbulent fragmentation toward an IMF, Salpeter-like mass spectrum.}

   \titlerunning{Kinematics of the Ophiuchus protocluster condensations}
   \authorrunning{Andr\'e et al.}
   \maketitle
%
%________________________________________________________________

%%%%%%%%%%%%%%%%%%
%%%%
\section{Introduction}
\label{intro}

While most stars are believed to form in clusters \citep[e.g.][]{Adams01,Lada03},
our present theoretical understanding of the star 
formation process is essentially limited to isolated dense cores and 
protostars \citep[e.g.][]{Shu87,Shu04}. 
Studying the formation and detailed properties of prestellar condensations 
in cluster-forming clouds is thus of prime importance if we are to 
explain the origin of the stellar initial mass function (IMF).
Some progress has recently been made in this area 
%\citep[cf.][; Motte \& Andr\'e 2001; and Ward-Thompson et al. 2007 for reviews]{Andre00}. 
\citep[cf.][ for reviews]{Andre00,MA01,Ward06}.

\begin{figure} [!ht]
\vspace*{-0.5 cm}
\centerline{\resizebox{1.0\hsize}{!}{\includegraphics[angle=270]{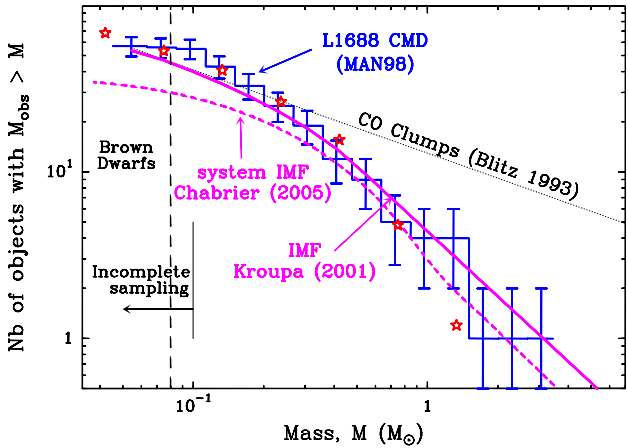}}}
\caption[Mass spectrum of $\rho$~Oph starless condensations]
{Cumulative mass distribution of the 57 starless condensations identified by \citeauthor*{Motte98} 
in the 
%$\rho$~Oph 
 L1688 protocluster (histogram with error bars). 
For comparison, the solid curve shows the shape of the field star IMF \citep[e.g.][]{Kroupa01},   
while the dashed curve corresponds to the IMF of multiple systems 
\citep[e.g.][]{Chabrier03,Chabrier05}.  The star markers represent the mass function of 
%$\rho$~Oph 
(primary) pre-main sequence objects in  L1688  
as derived from the ISOCAM mid-IR survey of \citet{Bontemps01}. 
The dotted line shows a $N(>M) \propto M^{-0.6}$ power-law 
distribution corresponding to the typical mass spectrum found 
for CO clumps \citep[see][]{Blitz93,Kramer98}.
Note the flattening of the mass distributions below $\sim 0.4\, M_\odot $ 
and the apparent excess of starless condensations over stellar systems 
at the low-mass end.}
\label{fig_preimf}
\end{figure}

On the observational side, recent ground-based (sub)-millimeter continuum
surveys of a few nearby cluster-forming clouds such as 
the L1688 clump in Ophiuchus (also known as the $\rho$~Ophiuchi main cloud; $d \sim$ 150~pc), 
the Serpens central clump ($d \sim 300$~pc), or the NGC~2068/2071 clumps 
in Orion~B ($d \sim 400$~pc) have uncovered `complete' 
samples of prestellar condensations whose associated mass distributions
resemble the stellar IMF 
\citep[e.g. Motte, Andr\'e, \& Neri 1998 -- \citeauthor*{Motte98};][; and references therein]{Testi98,Johnstone00,Motte01,Bontemps01}. 
In particular, this is the case for the population of 57 starless 
condensations identified by \citeauthor*{Motte98} in their 1.2~mm continuum mosaic of L1688 with 
the MPIfR bolometer array (MAMBO) on the IRAM 30m telescope.
These Ophiuchus condensations, which were identified using a multi-resolution 
analysis equivalent to a wavelet decomposition \citep[cf.][]{Starck98, Motte03}, are seen 
{\it on the same spatial scales as protostellar envelopes} (i.e., $\sim $~2300--4500~AU or 
$\sim 15\arcsec - 30\arcsec $ in L1688). 
Their mass spectrum is consistent with the \citet{Salpeter55} power-law IMF 
at the high-mass end  and shows a tentative break at $\sim 0.4\, M_\odot $ 
(see Fig.~\ref{fig_preimf}). 
This break is reminiscent of the flattening observed in the IMF of field
stars below $0.5\, M_\odot $ \citep[e.g.][]{Kroupa01,Chabrier03}, 
also present in the mass function of L1688 pre-main sequence objects 
\citep[][]{Luhman00,Bontemps01}.  
If real, the break occurs at a mass comparable to the typical Jeans mass 
in the dense ($n\h2 \sim 10^5\, \cm 3$) DCO$^+$ cores of the central Ophiuchus
molecular cloud \citep[cf.][]{Loren90}. 
The results of \citeauthor*{Motte98} in L1688 were essentially confirmed by 
independent (sub)-millimeter dust continuum surveys of the same region 
with SCUBA on JCMT \citep[][]{Johnstone00} and SIMBA on SEST 
\citep[][]{Stanke06}. 

Such a close resemblance of their mass spectrum to the IMF, 
in both shape and mass scale, suggests that the starless 
condensations identified by \citeauthor*{Motte98} 
%in the $\rho$~Oph protocluster 
are about to form stars on a one-to-one or perhaps one-to-two basis, 
with a high local efficiency, i.e., $M_\star /M_{pre} \simgt 50\%$. 
This strongly supports scenarios according to which  
the bulk of the IMF is at least partly determined by pre-collapse cloud 
fragmentation \citep[e.g.][]{Larson85,Larson05,Elmegreen97,Padoan02}.
The problem of the origin of the IMF may thus partly 
reduce to a good understanding of the processes responsible for 
the formation and evolution of prestellar condensations.
%within cluster-forming clouds.
Additional processes are likely to be required, however, 
to account for the formation of binary/multiple systems
and fully explain the low-mass ($M < 0.3\, M_\odot $) end of the IMF.
Indeed, while most young stars are observed to be in close multiple systems 
\citep[e.g.][]{Duchene04}, the 1.2mm continuum survey of \citeauthor*{Motte98}  
did not have enough spatial resolution to probe multiplicity within the 
%$\rho$~Oph 
 L1688 condensations. Furthermore, multiple systems are believed to 
form {\it after} the prestellar stage by subsequent dynamical fragmentation 
during the collapse phase, close to the time of protostar formation 
%(e.g. Goodwin et al. 2007). 
\citep[e.g.][]{Goodwin06}. 
% Ref Goodwin et al. 2006 ?
One would thus expect the masses of the  Ophiuchus 
%$\rho$~Oph 
prestellar condensations to be more directly related to the masses 
of multiple systems than to the masses of individual stars. 
Surprisingly, the shape of the condensation mass spectrum agrees better with 
the IMF of individual field stars (solid curve in Fig.~1) than with the 
IMF of multiple systems (dashed curve in Fig.~1).
Clearly, the link between the condensation mass spectrum and the IMF is less 
robust at the low-mass end than at the high-mass end.

On the theoretical side, two main scenarios have been proposed 
for clustered star formation in turbulent molecular clouds.
In the first scenario, the distribution of stellar masses 
is primarily determined by {\it gravoturbulent cloud fragmentation at the 
prestellar stage}. 
Briefly, self-gravitating condensations 
form as turbulence-generated density fluctuations 
\citep[e.g.][]{Klessen00,Padoan02}, then decouple from 
their turbulent environment through the dissipation of MHD waves on small
scales \citep[e.g.][]{Nakano98,Myers98}, 
and eventually collapse with little interaction with their 
surroundings. Such protocluster condensations 
are local minima of turbulence (traced by narrow linewidths) and have 
small relative motions with respect to one another and to the surrounding 
gas. A given star is entirely formed from (a fraction of) the gas that 
was initially present in the corresponding prestellar condensation. 
Thus, in this scenario, the IMF derives directly from the condensation mass 
distribution (CMD) \citep[][]{Padoan02}, 
which provides a simple explanation for the observed similarity between the CMD and the IMF 
(e.g. \citeauthor*{Motte98} and Fig.~\ref{fig_preimf}).\\ 
By contrast, in the second scenario, the distribution of stellar masses 
results entirely from the {\it dynamics of the parent protocluster} 
\citep[e.g.][]{Bonnell98,Bonnell01b}. 
Here, a protocluster is initially made up of gas and protostellar seeds.  
These protostellar seeds result from turbulence-generated cloud structure 
like in the first scenario, but their initial masses are unrelated to final stellar masses.
The seeds travel in the gravitational potential well of the system 
and are characterized by a large, essentially virial velocity dispersion. 
They accrete mass competitively as they execute several orbits through 
the protocluster.
The seed trajectories within the protocluster are highly stochastic in nature 
and feature close encounters, merging and/or dynamical ejections.
In this alternative scenario, competitive accretion and 
dynamical interactions between individual protocluster members play a 
dominant role in shaping the resulting IMF at the protostellar (Class~0/Class~I) 
stage \citep[e.g.][]{Bate03}. 
Furthermore, most of the final mass of a given star comes from gas 
that was initially not gravitationally bound to the corresponding protostellar 
seed(s) \citep[cf.][]{Bonnell04}.

In an effort to discriminate between these two broad pictures for the origin of 
the IMF and further constrain the nature of the starless condensations identified 
by \citeauthor*{Motte98}, we carried out detailed molecular line observations of the 
%$\rho$~Oph 
central Ophiuchus protocluster with the IRAM 30m telescope.
The present paper describes the results of these line observations and 
discusses them in the context of the above-mentioned theoretical scenarios 
for clustered star formation.\\
The layout of the paper is as follows. Section~\ref{obs_set} provides
observational details. Section~\ref{obs_ana} presents the results of our 
line mapping observations. Section~\ref{oph_kin} analyzes the constraints 
set by these observations on the kinematics of the 
%$\rho$~Oph 
 L1688 protocluster. 
We discuss the implications of our results for our understanding of cluster-forming 
clouds in Sect.~\ref{dis}. Our conclusions are summarized in Sect.~\ref{summary}.

\begin{table*}
\centering
 \caption{Adopted line rest frequencies and telescope efficiencies.}
 \label{tab_freq_oph}
 \vspace*{1.ex}
 \centering
 \footnotesize
 \begin{tabular}{lcclcccc}
 \hline
 \hline
 \multicolumn{1}{c}{Transition} & \multicolumn{1}{c}{Frequency$^a$} & \multicolumn{1}{c}{$\sigma_v^b$} & \multicolumn{1}{c}{Ref.$^c$} & \multicolumn{1}{c}{HPBW$^d$} & F$_{\mathrm{eff}}^e$ & B$_{\mathrm{eff}}^f$ & B$_{\mathrm{eff}}$ \\
 \multicolumn{1}{c}{} & \multicolumn{1}{c}{(MHz)} & \multicolumn{1}{c}{(km~s$^{-1}$)} & \multicolumn{1}{c}{} & \multicolumn{1}{c}{($\arcsec$)} & \multicolumn{1}{c}{} & \multicolumn{1}{c}{(2000)} & \multicolumn{1}{c}{(1998)} \\*[0.4ex]
 \hline
 H$^{13}$CO$^+$(1-0)        & \,\,\,\,\,\,86754.294(30) & 0.10  & (1) & 28.4 & 0.92 & 0.77 & \\
 N$_2$H$^+$(101-012)        & \,\,\,93176.265(7)        & 0.023 & (2) & 26.4 & 0.92 & 0.77 & 0.73 \\
 C$^{34}$S(2-1)             & \,\,\,96412.952(1)        & 0.003 & (3) & 25.5 & 0.92 & 0.77 & 0.73 \\
 CS(2-1)                    & \,\,\,97980.953(1)        & 0.003 & (3) & 25.1 & 0.92 & 0.80 & 0.73 \\
 H$_2$CO(2$_{12}$-1$_{11}$) & 140839.518(7)             & 0.015 & (1) & 17.5 & 0.90 & 0.65 & 0.54 \\
 DCO$^+$(2-1)               & \,\,\,144077.319(50)      & 0.10  & (4) & 17.1 & 0.90 & 0.65 & \\
 CS(3-2)                    & 146969.026(1)             & 0.002 & (3) & 16.7 & 0.90 & 0.65 & \\
% DCO$^+$(3-2)               & \,\,\,216112.604(50)      & 0.069 & (4) & 11.4 & 0.85 & 0.53 & \\
 HCO$^+$(3-2)               & \,\,\,267557.625(17)      & 0.019 & (1) &  9.2 & 0.85 & 0.49 & \\
 \hline
 \end{tabular}
 \vspace*{0.5ex}
 \begin{list}{}{}
  \item[$(a)$]{The frequency uncertainty in units of the last digit is given in parentheses.}
  \item[$(b)$]{Frequency uncertainty converted to velocity units.}
  \item[$(c)$]{Reference for the frequency: (1) \citet{Lovas92}, (2) \citet{Caselli95}, 
    (3) \citet*{Gottlieb03}, (4) \citet{Pickett98}.}
  \item[$(d)$]{Half-power beamwidth (HPBW) of the IRAM 30m telescope.}
  \item[$(e)$]{The forward efficiency was 0.95 for N$_2$H$^+$(1-0) in 2005 (OTF map in Oph~B1).}
  \item[$(f)$]{The main-beam efficiency was 0.77 for N$_2$H$^+$(1-0) in 2005.}
 \end{list}
 \normalsize 
\end{table*}

%%%%%%%%%%%%%%%%%%
%%%%

\begin{figure*} [!ht]
\centerline{\resizebox{0.85\hsize}{!}{\includegraphics[angle=0]{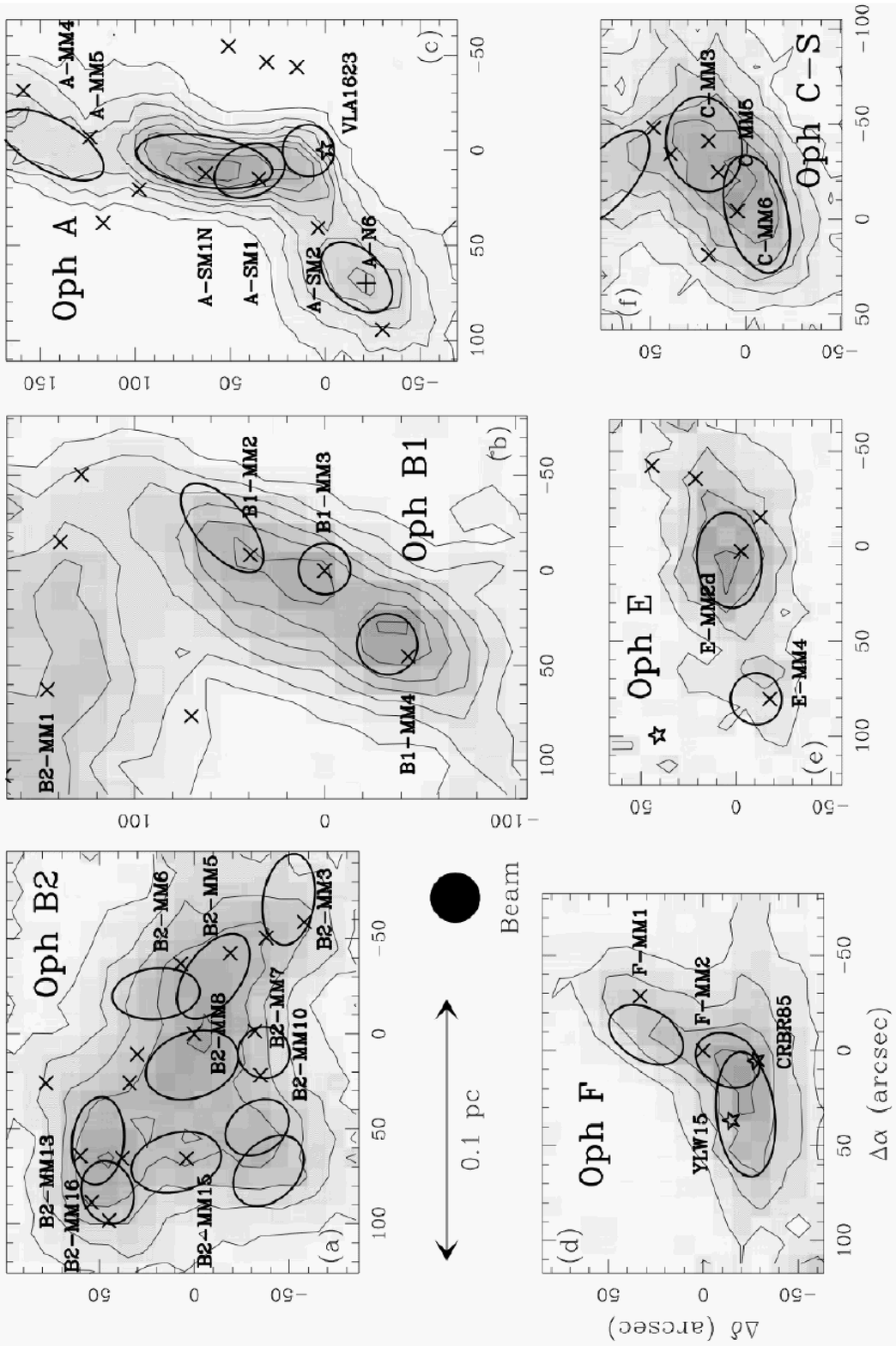}}}

%\vspace*{-0.3 cm}
\caption[N$_2$H\/$^+$(1-0) integrated intensity maps of the DCO$^+$ cores Oph~A, B1, B2, C, E, F]
{Maps of the N$_2$H\/$^+$(1-0) intensity, integrated over the seven components of the hyperfine 
multiplet, taken in the OTF mode with the IRAM 30~m telescope toward 
the DCO$^+$ cores Oph~B2 (a), Oph~B1 (b), Oph~A (c),  Oph~F (d), Oph~E (e), and Oph~C-S (f). 
%the DCO$^+$ cores Oph~A (a), Oph~B1 (b), Oph~B2 (c),  Oph~C-S (d), Oph~E (e), and Oph~F (f).
The (0,0) offsets correspond to the J2000 equatorial positions 
($16^{\mathrm{h}}27^{\mathrm{m}}27\fs96$,$-24^\circ27\arcmin06.9\arcsec$), 
($16^{\mathrm{h}}27^{\mathrm{m}}12\fs41$,$-24^\circ29\arcmin58.0\arcsec$), 
($16^{\mathrm{h}}26^{\mathrm{m}}26\fs45$,$-24^\circ24\arcmin30.8\arcsec$), 
($16^{\mathrm{h}}27^{\mathrm{m}}24\fs25$,$-24^\circ40\arcmin35.2\arcsec$), 
($16^{\mathrm{h}}27^{\mathrm{m}}04\fs70$,$-24^\circ39\arcmin12.5\arcsec$), and
($16^{\mathrm{h}}27^{\mathrm{m}}01\fs91$,$-24^\circ34\arcmin40.7\arcsec$), 
respectively.
The contours go from 2 to 10, 1 to 7, 2 to 16, 1.5 to 6, 0.75 to 3,  and 1 to 6 K~km~s$^{-1}$
by steps of 2, 1, 2, 1.5, 0.75, and 1 K~km~s$^{-1}$, respectively (in T$_{\mathrm{a}}^\star$ scale). 
The angular resolution (HPBW) is shown as a black filled circle.
The crosses mark the positions of the starless condensations identified by 
\citeauthor*{Motte98} in the dust continuum at 1.2~mm, the plus symbol 
the position of the N$_2$H$^+$ peak N6 discussed by 
\citet{DiFrancesco04}, and the star symbols the positions of Class~0 or 
Class~I protostars.
The ellipses show the locations of the clumps identified with {\it Gaussclumps}
in the corresponding N$_2$H\/$^+$(101-012) background-subtracted data cubes 
(see Sect.~\ref{n2h_det} and Table~\ref{t:gaussclumps_results}).
}
\label{fig_oph_mapn2h+}
\end{figure*}

\begin{figure*} [!ht]
%\centerline{\resizebox{0.9\hsize}{!}{\includegraphics[angle=270]{pandre_oph_line_fig3.ps}}}
\centerline{\resizebox{0.85\hsize}{!}{\includegraphics[angle=270]{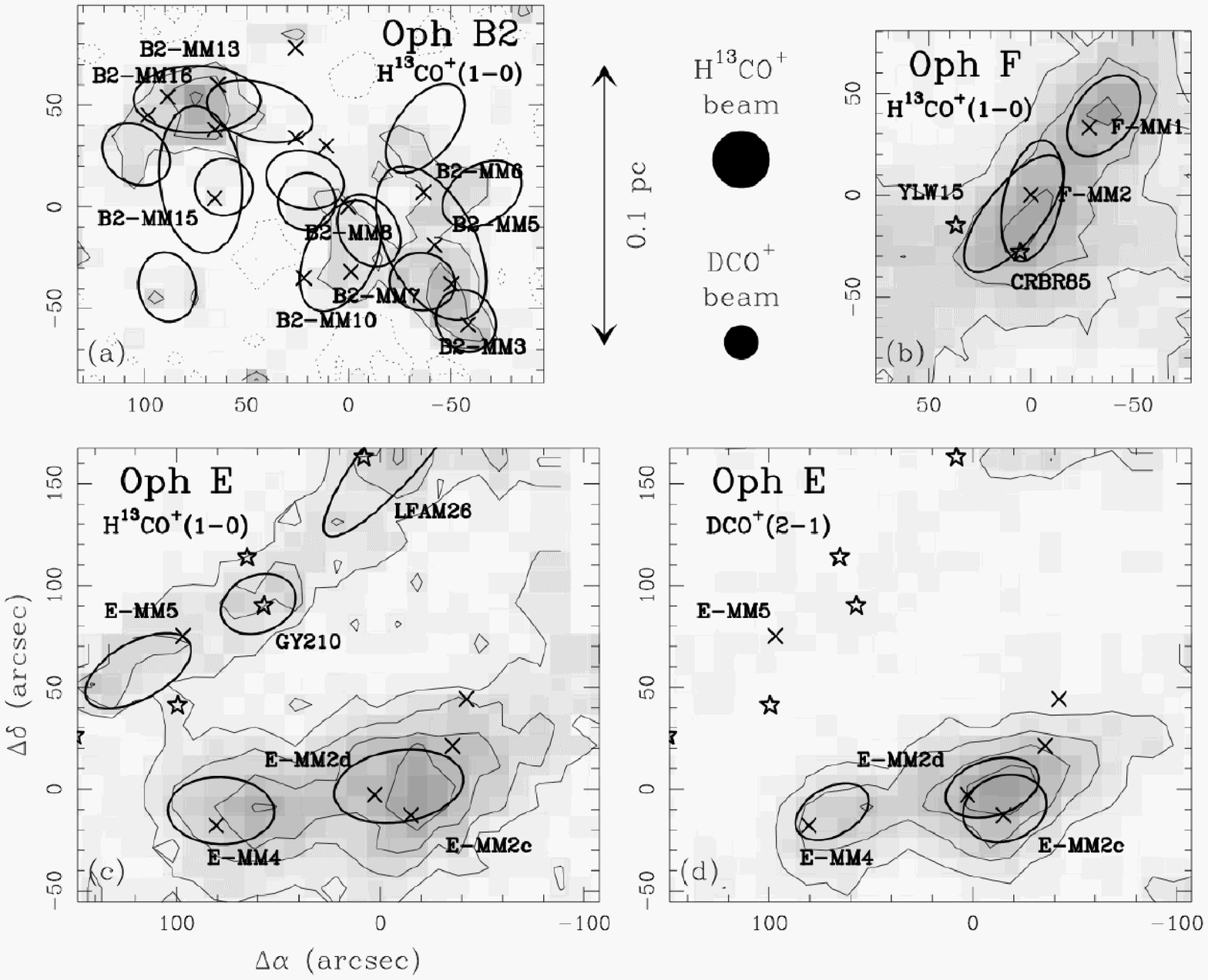}}}
\vspace*{-0.3 cm}
\caption[H$^{13}$CO$^+$(1-0)/DCO$^+$(2-1) integrated intensity maps of the DCO$^+$ cores Oph~B2, E, F]
{H$^{13}$CO$^+$(1-0) or DCO$^+$(2-1) integrated intensity maps taken in the OTF mode with the IRAM 30~m telescope 
toward the DCO$^+$ cores Oph~B2 (a), Oph~F (b), and Oph~E (c and d). 
The contours go from 0.3 to 0.9, 0.5 to 1.5, 0.25 to 1, and 0.6 to 2.4 K~km~s$^{-1}$ 
by steps of 0.3, 0.5, 0.25, 0.6 K~km~s$^{-1}$, respectively (in T$_{\mathrm{a}}^\star$ scale). 
In (a), the negative contours are -0.6 and -0.3 K~km~s$^{-1}$.
To emphasize small-scale structure, 
large-scale ($> 40 \arcsec $) emission was subtracted from the H$^{13}$CO$^+$(1-0) map of Oph~B2 shown in (a)
(see Sect.~\ref{n2h_det}).
The angular resolution (HPBW) is shown as a black filled circle.
The crosses mark the positions of the starless condensations identified by \citeauthor*{Motte98} in the dust
continuum at 1.2~mm, 
and the star symbols the positions of Class~0 or Class~I protostars.
The ellipses show the locations of the clumps identified with {\it Gaussclumps}
in the corresponding background-subtracted data cubes 
(see Sect.~\ref{n2h_det}).
%and Table~\ref{t:veldisp_results}).
}
\label{fig_oph_maph13co+}
\end{figure*}

\section{Observations}
\label{obs_set}

We used the IRAM 30m telescope at Pico Veleta, Spain, in June 1998, July 2000, and June 2005 
to carry out millimeter line observations of the DCO$^+$ cores and 
1.2~mm continuum condensations  of L1688  
in the following molecular transitions:\\
N$_2$H$^+$(1-0), H$^{13}$CO$^+$(1-0), 
CS(2-1), C$^{34}$S(2-1) at 3~mm, 
CS(3-2), H$_2$CO(2$_{12}$-1$_{11}$), DCO$^+$(2-1) at 2~mm, and HCO$^+$(3-2)
at 1.1~mm.
Our adopted set of rest line frequencies is given in Table \ref{tab_freq_oph}. 
The half-power beamwidth of the telescope was $\sim 26 ~\arcsec$, $\sim 17 ~\arcsec$, 
and $\sim 9 ~\arcsec$ at 3~mm, 2~mm, and 1.1~mm, respectively. 
We used four SIS heterodyne receivers simultaneously and an autocorrelation
spectrometer as backend, with a spectral resolution of 20--40~kHz at 3~mm and 2~mm, 
and 40~kHz at 1.1~mm. The corresponding velocity resolution ranged from 
0.04 to 0.07 km~s$^{-1}$ per channel, depending on the observed transition. 
All our observations were performed in single sideband mode, with sideband rejections 
of 0.01, 0.1 and 0.05 at 3~mm, 2~mm and 1.1~mm, respectively. 
The resulting calibration uncertainty is $\sim 10 \%$. 
The forward and beam efficiencies of the telescope 
used to convert antenna temperatures $T^*_\mathrm{A}$ into 
main beam temperatures $T_\mathrm{mb}$ are listed in Table~\ref{tab_freq_oph}.
The telescope pointing was checked every $\sim 2$ hours 
on NRAO~530 and/or 1514-241 and found to be accurate to 
$\sim 4\arcsec$ (rms).
The telescope focus was optimized on 3C273 and NRAO~530 every $\sim 2$ hours on average.
Single-point line observations were performed in the position switching mode, 
while extensive mapping was performed in the ``on-the-fly'' (OTF) mode.
All of the data were reduced with the CLASS software package\footnote{see  
http://www.iram.fr/IRAMFR/GILDAS.}.

%%%%%%%%%%%%%%%%%%
%%
\section{Molecular line mapping results and analysis}
\label{obs_ana}

%%%
%
\subsection{Detections of protocluster condensations in N$_2$H$^+$(1--0) and other tracers}
\label{n2h_det}

The N$_2$H$^+$(1-0) integrated intensity maps\footnote{Our N$_2$H$^+$(1-0) data cubes are available in FITS format at the CDS.} 
we took toward the six main DCO$^+$ cores of the Ophiuchus 
%$\rho$ Ophiuchi
central cloud are shown in Fig.~\ref{fig_oph_mapn2h+}. The positions of the starless 1.2~mm continuum 
condensations found by \citeauthor*{Motte98} are marked by crosses. 
N$_2$H$^+$(1-0) emission was found to be present toward most of these condensations. 
Altogether, we performed N$_2$H$^+$(1-0) observations toward 48 
of the 57 compact starless condensations\footnote{\citeauthor*{Motte98} counted 58 compact starless condensations but one 
of these -- E-MM3 -- was subsequently shown to be an edge-on T Tauri disk as opposed to a prestellar object \citep[e.g.][]{Brandner00}.} 
identified by \citeauthor*{Motte98}
and detected line emission for 41 of them, either through position-switch integrations (rms $\sim 0.1$~K in $T_A^*$ units) 
or in OTF maps (rms $\sim $~0.1--0.2~K, depending on core).  
However, the $\sim 26\arcsec$ angular resolution of our N$_2$H$^+$ 
observations is not always sufficient to clearly distinguish between line emission from the condensations 
themselves and emission from the parent cloud/DCO$^+$ core. In order to estimate the background N$_2$H$^+$ 
emission from the parent core, we used our OTF maps to produce a smoothed N$_2$H$^+$(1-0) image of the 
local background  toward each object. In practice, we used the multi-resolution wavelet analysis software of 
\citet{Starck98} to decompose our OTF N$_2$H$^+$(1-0) data cubes into two 
wavelet  ``views'', corresponding to small ($\sim  $~20-40\arcsec) and large ($\simgt $~40\arcsec) spatial scales, respectively.
The background emission was estimated from the ``large-scale'' view, while an estimate of the background-subtracted emission 
from the condensations was provided by the ``small-scale'' view. 
 Such a multi-resolution decomposition is similar to the analysis performed by \citeauthor*{Motte98} to separate compact 
condensations and protostellar envelopes seen on angular scales $\simlt  $~30-60\arcsec ~from more extended 
($\simgt 1\arcmin $) cloud structure in their 1.2~mm dust continuum image. The fact that most of the condensations detected 
in OTF maps remained positively detected after background subtraction demonstrates that our N$_2$H$^+$ observations 
successfully probed the condensations themselves and not only the dense environment of the parent clump/DCO$^+$ cores.

Additional H$^{13}$CO$^+$(1-0) and DCO$^+$(2-1) OTF maps were obtained toward some of the cores 
(cf. Fig.~\ref{fig_oph_maph13co+}), which were background-subtracted and analyzed in a similar fashion.
\begin{table*} 
 \centering 
 \caption{Results of Gaussian hyperfine fits to the N$_2$H$^+$(1-0) multiplet.} 
 \label{t:fithfs_results} 
\vspace*{0.0ex} 
 {\scriptsize 
 \begin{tabular}{lcccccccccccccc} 
 \hline\hline 
 & & & \multicolumn{6}{c}{Prior to background subtraction} &  \multicolumn{6}{c}{After background subtraction}  \\  
 Source & \multicolumn{2}{c}{Coordinates} & $\sigma^a$ & \hspace*{-2ex} S$/$N$^b$ & \hspace*{-2ex} $P_1^c$ & $V_{lsr}$ & FWHM & $\tau_{tot}^d$ & $\sigma^a$ &  \hspace*{-2ex} S$/$N$^b$ &  \hspace*{-2ex} $P_1^c$ & $V_{lsr}$ & FWHM & $\tau_{tot}^d$ \\  
  & \hspace*{-2ex} $\alpha_{2000}$ & \hspace*{-2ex} $\delta_{2000}$ & {\scriptsize (mK)} & &  \hspace*{-2ex} {\scriptsize (K)} & {\scriptsize (km~s$^{-1}$)} & {\scriptsize (km~s$^{-1}$)} & & {\scriptsize (mK)} & &  \hspace*{-2ex} {\scriptsize (K)} & {\scriptsize (km~s$^{-1}$)} & {\scriptsize (km~s$^{-1}$)} \\  
  & 16$^{\mbox{h}}$ & -24$^\circ$ & & & & & & & & & & & & \\  
(1) & (2) & (3) & (4) & (5) & (6) & (7) & (8) & (9) & (10) & (11) & (12) & (13) & (14) & (15) \\  
 \hline 
A3-MM1  &  \hspace*{-2ex}  26:09.7   &  \hspace*{-2ex}  23:06  &          79  & \hspace*{-2ex}           7  & \hspace*{-2ex}     1.6(2)  &       3.243(26)  &       0.877(59)  &         0.2(24)  &  & & & & & \\  
A-MM4  &  \hspace*{-2ex}  26:24.1   &  \hspace*{-2ex}  21:52  &          65  & \hspace*{-2ex}          45  & \hspace*{-2ex}    15.8(1)  &       3.243(2)  &       0.445(2)  &         2.6(1)  &         67  &  \hspace*{-2ex}          17  &  \hspace*{-2ex}     4.5(5)  &       3.324(8)  &       0.378(18)  &         1.5(13)  \\  
A-MM5  &  \hspace*{-2ex}  26:25.9   &  \hspace*{-2ex}  22:27  &          62  & \hspace*{-2ex}          53  & \hspace*{-2ex}    15.3(5)  &       3.171(2)  &       0.381(6)  &         1.9(3)  &         64  &  \hspace*{-2ex}          15  &  \hspace*{-2ex}     9.8(16)  &       3.146(5)  &       0.183(10)  &        10.6(29)  \\  
SM1N  &  \hspace*{-2ex}  26:27.3   &  \hspace*{-2ex}  23:28  &          60  & \hspace*{-2ex}          98  & \hspace*{-2ex}    52.2(8)  &       3.573(1)  &       0.487(3)  &         8.6(2)  &         57  &  \hspace*{-2ex}          47  &  \hspace*{-2ex}    29.0(9)  &       3.587(3)  &       0.436(6)  &        11.4(5)  \\  
SM1  &  \hspace*{-2ex}  26:27.5   &  \hspace*{-2ex}  23:56  &          43  & \hspace*{-2ex}         125  & \hspace*{-2ex}    35.9(0)  &       3.645(1)  &       0.597(1)  &         5.7(0)  &         43  &  \hspace*{-2ex}          56  &  \hspace*{-2ex}    15.2(4)  &       3.682(3)  &       0.640(7)  &         6.1(3)  \\  
A-MM6  &  \hspace*{-2ex}  26:27.9   &  \hspace*{-2ex}  22:53  &          64  & \hspace*{-2ex}          20  & \hspace*{-2ex}     5.0(3)  &       3.353(6)  &       0.694(16)  &         1.4(5)  &  & & & & & \\  
SM2  &  \hspace*{-2ex}  26:29.5   &  \hspace*{-2ex}  24:27  &          52  & \hspace*{-2ex}          68  & \hspace*{-2ex}    19.8(3)  &       3.509(1)  &       0.485(3)  &         3.6(2)  &         51  &  \hspace*{-2ex}          19  &  \hspace*{-2ex}     5.8(5)  &       3.519(6)  &       0.386(13)  &         4.9(12)  \\  
A-MM8  &  \hspace*{-2ex}  26:33.4   &  \hspace*{-2ex}  25:01  &          65  & \hspace*{-2ex}          51  & \hspace*{-2ex}    17.3(4)  &       3.505(1)  &       0.384(4)  &         2.7(3)  &         64  &  \hspace*{-2ex}          16  &  \hspace*{-2ex}     3.9(1)  &       3.571(6)  &       0.332(12)  &         0.1(6)  \\  
A-S  &  \hspace*{-2ex}  26:43.1   &  \hspace*{-2ex}  25:42  &          81  & \hspace*{-2ex}           7  & \hspace*{-2ex}     2.9(6)  &       3.710(7)  &       0.236(20)  &         2.2(22)  &  & & & & & \\  
VLA1623  &  \hspace*{-2ex}  26:26.5   &  \hspace*{-2ex}  24:31  &          37  & \hspace*{-2ex}          75  & \hspace*{-2ex}    17.1(2)  &       3.685(1)  &       0.561(3)  &         4.4(1)  &         37  &  \hspace*{-2ex}          30  &  \hspace*{-2ex}     6.9(3)  &       3.749(4)  &       0.532(8)  &         4.2(5)  \\  
\hline 
B1-MM1  &  \hspace*{-2ex}  27:08.7   &  \hspace*{-2ex}  27:50  &          91  & \hspace*{-2ex}          10  & \hspace*{-2ex}     3.1(1)  &       4.069(8)  &       0.383(21)  &         0.1(2)  &  & & & & & \\  
B1-MM2  &  \hspace*{-2ex}  27:11.8   &  \hspace*{-2ex}  29:19  &         101  & \hspace*{-2ex}          26  & \hspace*{-2ex}     5.4(5)  &       3.387(19)  &       0.826(39)  &         2.2(7)  &        100  &  \hspace*{-2ex}          11  &  \hspace*{-2ex}     1.9(2)  &       3.430(42)  &       0.981(88)  &         0.1(203)  \\  
 & & & & &    10.2(8)  &       4.050(4)  &       0.287(11)  &         3.2(9)  &  & &     4.8(10)  &       4.060(8)  &       0.217(26)  &         5.1(28)  \\  
B1-MM3  &  \hspace*{-2ex}  27:12.4   &  \hspace*{-2ex}  29:58  &          67  & \hspace*{-2ex}          38  & \hspace*{-2ex}     9.7(8)  &       3.287(4)  &       0.280(11)  &        11.1(15)  &  & & & & & \\  
 & & & & &    14.6(4)  &       3.821(2)  &       0.340(5)  &         3.3(4)  &         70  &  \hspace*{-2ex}          19  &  \hspace*{-2ex}    16.9(22)  &       3.780(3)  &       0.185(7)  &        13.7(27)  \\  
B1-MM4  &  \hspace*{-2ex}  27:15.7   &  \hspace*{-2ex}  30:42  &         216  & \hspace*{-2ex}          15  & \hspace*{-2ex}    34.8(68)  &       3.551(5)  &       0.158(11)  &        23.5(54)  &        221  &  \hspace*{-2ex}           8  &  \hspace*{-2ex}    14.8(14)  &       3.762(27)  &       0.451(37)  &        29.8(37)  \\  
 & & & & &    14.6(12)  &       3.956(5)  &       0.341(14)  &         2.2(10)  &  & &     4.3(7)  &       3.967(11)  &       0.205(22)  &         0.3(13)  \\  
\hline 
B1B2-MM1  &  \hspace*{-2ex}  27:11.3   &  \hspace*{-2ex}  27:39  &         130  & \hspace*{-2ex}          16  & \hspace*{-2ex}     7.6(7)  &       4.070(6)  &       0.384(16)  &         0.2(12)  &        128  &  \hspace*{-2ex}           5  &  \hspace*{-2ex}     3.0(11)  &       4.065(12)  &       0.217(39)  &         1.5(37)  \\  
B1B2-MM2$^e$  &  \hspace*{-2ex}  27:18.0   &  \hspace*{-2ex}  28:48  &         126  & \hspace*{-2ex}           6  & \hspace*{-2ex}     2.7(6)  &       3.957(25)  &       0.556(63)  &         2.4(25)  &  & & & & & \\  
\hline 
B2-MM1  &  \hspace*{-2ex}  27:17.0   &  \hspace*{-2ex}  27:32  &         147  & \hspace*{-2ex}          16  & \hspace*{-2ex}    12.6(11)  &       4.049(6)  &       0.380(14)  &         4.5(11)  &        144  &  \hspace*{-2ex}           4  &  \hspace*{-2ex}     6.2(37)  &       3.984(15)  &       0.184(42)  &        15.1(114)  \\  
B2-MM2  &  \hspace*{-2ex}  27:20.3   &  \hspace*{-2ex}  27:08  &          99  & \hspace*{-2ex}          18  & \hspace*{-2ex}     7.5(2)  &       3.947(19)  &       0.757(34)  &         3.7(4)  &  & & & & & \\  
 & & & & &     6.7(7)  &       4.344(7)  &       0.238(18)  &        10.3(0)  &  & & & & & \\  
B2-MM3  &  \hspace*{-2ex}  27:23.7   &  \hspace*{-2ex}  28:05  &         130  & \hspace*{-2ex}          14  & \hspace*{-2ex}     8.4(2)  &       3.785(9)  &       0.531(11)  &         3.0(1)  &  & & & & & \\  
 & & & & &     4.2(9)  &       4.376(4)  &       0.305(30)  &         2.1(23)  &  & & & & & \\  
B2-MM4  &  \hspace*{-2ex}  27:24.3   &  \hspace*{-2ex}  27:45  &         262  & \hspace*{-2ex}          14  & \hspace*{-2ex}    21.7(19)  &       3.718(7)  &       0.393(17)  &         6.4(13)  &        259  &  \hspace*{-2ex}           5  &  \hspace*{-2ex}    14.9(46)  &       3.678(12)  &       0.262(25)  &        16.3(66)  \\  
 & & & & &     4.6(7)  &       4.358(15)  &       0.345(38)  &         0.1(22)  &  & & & & & \\  
B2-MM5  &  \hspace*{-2ex}  27:24.9   &  \hspace*{-2ex}  27:26  &         246  & \hspace*{-2ex}          13  & \hspace*{-2ex}    15.8(4)  &       3.693(1)  &       0.497(19)  &         2.9(0)  &        238  &  \hspace*{-2ex}           5  &  \hspace*{-2ex}     7.9(21)  &       3.644(15)  &       0.305(28)  &         7.7(42)  \\  
 & & & & &     4.7(15)  &       4.388(62)  &       0.597(97)  &         8.0(45)  &  & & & & & \\  
B2-MM6  &  \hspace*{-2ex}  27:25.3   &  \hspace*{-2ex}  27:00  &         240  & \hspace*{-2ex}          12  & \hspace*{-2ex}    15.6(13)  &       3.763(10)  &       0.699(26)  &         4.0(8)  &        248  &  \hspace*{-2ex}           4  &  \hspace*{-2ex}    10.1(28)  &       3.680(22)  &       0.424(39)  &        14.3(56)  \\  
B2-MM7  &  \hspace*{-2ex}  27:27.9   &  \hspace*{-2ex}  27:39  &         261  & \hspace*{-2ex}          10  & \hspace*{-2ex}     9.2(13)  &       4.214(21)  &       0.789(60)  &         2.0(11)  &  & & & & & \\  
B2-MM8  &  \hspace*{-2ex}  27:28.0   &  \hspace*{-2ex}  27:07  &         206  & \hspace*{-2ex}          19  & \hspace*{-2ex}    19.2(11)  &       4.144(6)  &       0.584(14)  &         3.0(6)  &        208  &  \hspace*{-2ex}           8  &  \hspace*{-2ex}     6.3(3)  &       4.186(10)  &       0.419(17)  &         0.1(5)  \\  
B2-MM9  &  \hspace*{-2ex}  27:28.8   &  \hspace*{-2ex}  26:37  &         240  & \hspace*{-2ex}          13  & \hspace*{-2ex}    11.3(12)  &       4.093(12)  &       0.650(33)  &         1.9(10)  &  & & & & & \\  
B2-MM10  &  \hspace*{-2ex}  27:29.6   &  \hspace*{-2ex}  27:42  &         162  & \hspace*{-2ex}          15  & \hspace*{-2ex}     8.9(8)  &       4.334(9)  &       0.569(22)  &         0.5(8)  &  & & & & & \\  
B2-MM11  &  \hspace*{-2ex}  27:29.8   &  \hspace*{-2ex}  25:49  &         145  & \hspace*{-2ex}          12  & \hspace*{-2ex}     4.2(2)  &       4.040(16)  &       1.018(51)  &         0.1(4)  &        146  &  \hspace*{-2ex}           4  &  \hspace*{-2ex}     4.4(26)  &       4.070(54)  &       0.633(134)  &        15.4(104)  \\  
B2-MM12  &  \hspace*{-2ex}  27:29.9   &  \hspace*{-2ex}  26:33  &         244  & \hspace*{-2ex}          14  & \hspace*{-2ex}    16.4(12)  &       4.093(8)  &       0.549(20)  &         3.5(9)  &        255  &  \hspace*{-2ex}           4  &  \hspace*{-2ex}     8.5(24)  &       4.097(12)  &       0.238(26)  &         7.7(44)  \\  
B2-MM13  &  \hspace*{-2ex}  27:32.7   &  \hspace*{-2ex}  26:07  &         263  & \hspace*{-2ex}          12  & \hspace*{-2ex}    14.1(14)  &       3.967(12)  &       0.637(29)  &         4.1(11)  &  & & & & & \\  
B2-MM14  &  \hspace*{-2ex}  27:32.8   &  \hspace*{-2ex}  26:29  &         258  & \hspace*{-2ex}          11  & \hspace*{-2ex}    14.9(14)  &       4.162(12)  &       0.771(32)  &         3.8(9)  &  & & & & & \\  
B2-MM15  &  \hspace*{-2ex}  27:32.8   &  \hspace*{-2ex}  27:03  &         267  & \hspace*{-2ex}          15  & \hspace*{-2ex}    31.6(24)  &       4.413(5)  &       0.356(11)  &         7.1(11)  &        272  &  \hspace*{-2ex}           6  &  \hspace*{-2ex}    14.9(36)  &       4.445(8)  &       0.221(17)  &         9.3(41)  \\  
B2-MM16  &  \hspace*{-2ex}  27:34.5   &  \hspace*{-2ex}  26:12  &          55  & \hspace*{-2ex}          54  & \hspace*{-2ex}    15.7(2)  &       4.041(2)  &       0.626(4)  &         3.5(2)  &         54  &  \hspace*{-2ex}          23  &  \hspace*{-2ex}     9.2(7)  &       4.074(7)  &       0.426(15)  &         6.8(10)  \\  
B2-MM17  &  \hspace*{-2ex}  27:35.2   &  \hspace*{-2ex}  26:21  &         250  & \hspace*{-2ex}          12  & \hspace*{-2ex}    13.1(14)  &       4.079(12)  &       0.622(28)  &         3.4(11)  &        257  &  \hspace*{-2ex}           5  &  \hspace*{-2ex}     4.2(14)  &       4.110(19)  &       0.320(45)  &         1.4(33)  \\  
\hline 
C-W$^e$  &  \hspace*{-2ex}  26:50.0   &  \hspace*{-2ex}  32:49  &         154  & \hspace*{-2ex}          12  & \hspace*{-2ex}    14.6(17)  &       3.578(4)  &       0.192(9)  &         7.2(18)  &  & & & & & \\  
C-N$^e$  &  \hspace*{-2ex}  26:57.2   &  \hspace*{-2ex}  31:39  &          93  & \hspace*{-2ex}          33  & \hspace*{-2ex}    36.3(16)  &       3.811(1)  &       0.211(3)  &        12.5(8)  &  & & & & & \\  
C-MM2  &  \hspace*{-2ex}  26:58.4   &  \hspace*{-2ex}  33:53  &         263  & \hspace*{-2ex}           8  & \hspace*{-2ex}    18.2(27)  &       3.879(12)  &       0.429(24)  &        11.5(26)  &  & & & & & \\  
C-MM3  &  \hspace*{-2ex}  26:58.9   &  \hspace*{-2ex}  34:22  &         281  & \hspace*{-2ex}           9  & \hspace*{-2ex}    35.2(46)  &       3.910(7)  &       0.307(15)  &        17.4(30)  &  & & & & & \\  
C-MM4  &  \hspace*{-2ex}  26:59.4   &  \hspace*{-2ex}  34:02  &         287  & \hspace*{-2ex}           9  & \hspace*{-2ex}    28.6(49)  &       3.852(9)  &       0.334(18)  &        15.7(38)  &  & & & & & \\  
C-MM5  &  \hspace*{-2ex}  27:00.1   &  \hspace*{-2ex}  34:27  &          62  & \hspace*{-2ex}          38  & \hspace*{-2ex}    38.5(12)  &       3.809(1)  &       0.314(3)  &        17.6(7)  &         62  &  \hspace*{-2ex}          16  &  \hspace*{-2ex}    22.1(7)  &       3.803(5)  &       0.238(6)  &        30.0(25)  \\  
C-MM6  &  \hspace*{-2ex}  27:01.6   &  \hspace*{-2ex}  34:37  &          61  & \hspace*{-2ex}          38  & \hspace*{-2ex}    37.1(22)  &       3.710(1)  &       0.327(6)  &        17.4(13)  &         61  &  \hspace*{-2ex}          18  &  \hspace*{-2ex}    24.8(26)  &       3.667(4)  &       0.249(7)  &        26.3(33)  \\  
C-MM7  &  \hspace*{-2ex}  27:03.3   &  \hspace*{-2ex}  34:22  &         294  & \hspace*{-2ex}           6  & \hspace*{-2ex}    34.1(15)  &       3.691(10)  &       0.294(11)  &        30.0(44)  &  & & & & & \\  
\hline 
E-MM1$^e$  &  \hspace*{-2ex}  26:57.7   &  \hspace*{-2ex}  36:56  &         153  & \hspace*{-2ex}           5  & \hspace*{-2ex}     2.1(2)  &       4.434(21)  &       0.360(39)  &         0.1(195)  &  & & & & & \\  
E-MM2d  &  \hspace*{-2ex}  27:04.9   &  \hspace*{-2ex}  39:15  &          61  & \hspace*{-2ex}          39  & \hspace*{-2ex}    13.5(4)  &       4.496(1)  &       0.287(4)  &         3.6(4)  &         62  &  \hspace*{-2ex}          21  &  \hspace*{-2ex}     9.0(7)  &       4.482(3)  &       0.253(8)  &         5.5(10)  \\  
E-MM4  &  \hspace*{-2ex}  27:10.6   &  \hspace*{-2ex}  39:30  &          41  & \hspace*{-2ex}          26  & \hspace*{-2ex}     5.1(3)  &       4.232(3)  &       0.331(7)  &         1.7(6)  &         41  &  \hspace*{-2ex}          16  &  \hspace*{-2ex}     4.8(6)  &       4.219(6)  &       0.254(13)  &         6.7(19)  \\  
\hline 
F-MM1  &  \hspace*{-2ex}  27:22.1   &  \hspace*{-2ex}  40:02  &          59  & \hspace*{-2ex}          26  & \hspace*{-2ex}     8.9(5)  &       4.726(5)  &       0.476(11)  &         4.6(7)  &         62  &  \hspace*{-2ex}          15  &  \hspace*{-2ex}     3.4(1)  &       4.789(6)  &       0.329(13)  &         0.1(9)  \\  
F-MM2  &  \hspace*{-2ex}  27:24.3   &  \hspace*{-2ex}  40:35  &          43  & \hspace*{-2ex}          53  & \hspace*{-2ex}    18.2(2)  &       4.133(0)  &       0.229(4)  &        10.8(0)  &         44  &  \hspace*{-2ex}          22  &  \hspace*{-2ex}     6.4(8)  &       4.148(4)  &       0.188(11)  &         9.6(23)  \\  
 & & & & &     7.7(2)  &       4.592(2)  &       0.354(6)  &         0.6(3)  &  & &     4.6(4)  &       4.584(1)  &       0.304(11)  &         2.0(10)  \\  
CRBR85  &  \hspace*{-2ex}  27:24.7   &  \hspace*{-2ex}  41:03  &         215  & \hspace*{-2ex}          13  & \hspace*{-2ex}    25.4(26)  &       4.067(4)  &       0.230(10)  &        10.2(17)  &        213  &  \hspace*{-2ex}           6  &  \hspace*{-2ex}    12.3(28)  &       4.032(8)  &       0.200(19)  &        11.1(41)  \\  
 & & & & &     3.3(3)  &       4.649(20)  &       0.402(44)  &         0.1(4)  &  & & & & & \\  
YLW15  &  \hspace*{-2ex}  27:27.0   &  \hspace*{-2ex}  40:50  &          37  & \hspace*{-2ex}          94  & \hspace*{-2ex}    29.5(4)  &       4.162(1)  &       0.258(2)  &         8.1(2)  &         36  &  \hspace*{-2ex}          54  &  \hspace*{-2ex}    19.0(10)  &       4.172(2)  &       0.212(4)  &        10.1(9)  \\  
\hline 
 \end{tabular}} 
 \begin{list}{}{} 
 \item[]{Notes: the numbers in parentheses indicate the uncertainty in units of the last digit.} 
\item[$(a)$]{$\sigma$ is the rms noise in T$_a^\star$ scale.} 
 \item[$(b)$]{S$/$N is the signal-to-noise ratio.} 
 \item[$(c)$]{The fitting function is $T_{a}^{\star }(v) = \frac{P_1}{\tau_{tot}} (1-e^{-\tau(v)})$. 
 In the optically thick case, $P_1 = T_{a}^{\star,peak} \times \tau_{tot}$}. 
\item[$(d)$]{$\tau_{tot}$ is the total optical depth of the 
 N$_2$H$^+$(1--0) multiplet. The optical depth of the isolated component 101--012 is $\frac{\tau_{tot}}{9}$.} 
 \item[$(e)$]{Composite starless clump (cf. \citeauthor*{Motte98}) not included 
 in the condensation mass distributions shown in Fig.~\ref{fig_preimf} and Fig.~\ref{fig_wcmd}.} 
\end{list} 
 \end{table*}

\begin{figure} [!ht]
%\vspace*{-0.5 cm}
\centerline{\resizebox{1.\hsize}{!}{\includegraphics[angle=270]{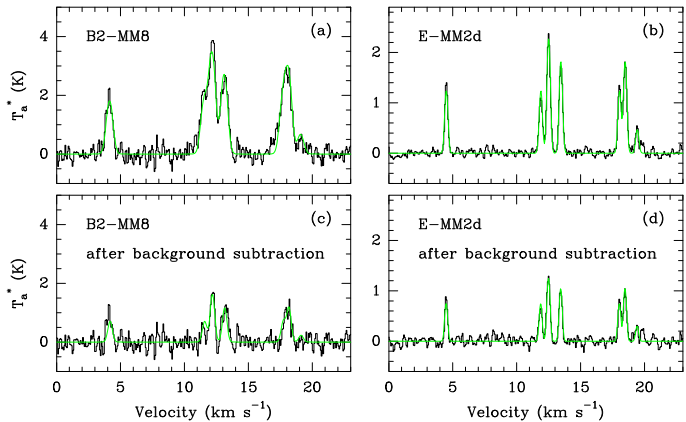}}}
%\vspace*{-0.3 cm}
\caption[Examples of N$_2$H\/$^+$(1-0) spectra]
{Examples of N$_2$H\/$^+$(1-0) spectra and Gaussian HFS fits obtained toward the condensations 
B2-MM8 (left) and E-MM2d (right). The spectra shown in the top row are those observed toward 
B2-MM8 (a) and E-MM2d (b) {\it prior to background subtraction}, while the spectra shown in the bottom
row are those obtained toward the same sources {\it after subtracting the local background} emission estimated
using a multi-resolution wavelet decomposition (see text).
}
\label{fig_oph_specn2h+}
\end{figure}

At the position of each target condensation/protostar, 
the seven hyperfine components of the N$_2$H$^+$(1-0) multiplet were fitted simultaneously using the 
Gaussian HFS (HyperFine Structure) fitting routine of the CLASS software package. 
This routine derives the line optical depth by assuming the same excitation temperature 
for all hyperfine components, and therefore yields an  estimate of the intrinsic linewidth (i.e., properly corrected 
for optical depth effects provided that the assumption is correct).
The results of these HFS fits, both before and after background subtraction, 
are given in Table~\ref{t:fithfs_results} 
for all the objects detected  in N$_2$H$^+$(1-0). 
(Note that, in some cases, the signal-to-noise ratio was not good enough to perform a significant HFS fit after background 
subtraction.)
The quoted error bars correspond to standard deviations ($1\sigma$) as estimated by the HFS routine 
of CLASS. 
Examples of N$_2$H$^+$(1-0) spectra and HFS fits are shown in Fig.~\ref{fig_oph_specn2h+}.
A single-component HFS fit failed and a two-component HFS fit was required  
for three condensations in Oph~B1, four condensations in Oph~B2, one condensation in Oph~F, as well as 
the protostar CRBR~85. 

\begin{table*} 
\centering 
\caption{Compact objects identified with \textit{Gaussclumps} in the background-subtracted N$_2$H$^+$(101-012)  data cubes.} 
\label{t:gaussclumps_results} 
\vspace*{0.5ex} 
\begin{tabular}{lcccccccccl} 
\hline\hline 
Object & $\Delta\alpha$ & $\Delta\delta$ & $v_0$ & $a_0$ & $\Delta x$ & $\Delta y$ & $\phi$ & FWHM & $\Delta v$ & Identification \\  
name & {\scriptsize ('')} & {\scriptsize ('')} & {\scriptsize (km~s$^{-1}$)} & {\scriptsize (K)} & {\scriptsize ('')} & {\scriptsize ('')} & {\scriptsize ($^\circ$)} & {\scriptsize ('')} & {\scriptsize (km~s$^{-1}$)} & \\  
(1) & (2) & (3) & (4) & (5) & (6) & (7) & (8) & (9) & (10) & (11) \\  
\hline 
A-1  &    6.0  &   63.6  &  3.52  &  2.16  &  27.00  &  72.50  &    81.2  &  35.8  &  0.49  &  A-SM1N  \\  
A-2  &   67.1  &  -16.6  &  3.42  &  1.85  &  28.40  &  44.70  &    46.6  &  33.9  &  0.37  &  A-N6$^a$  \\  
A-3  &   -2.6  &  144.1  &  3.14  &  1.01  &  27.00  &  60.90  &    61.5  &  34.9  &  0.34  &  $\sim$ A-MM5  \\  
A-6  &   10.9  &   39.7  &  3.90  &  0.65  &  27.00  &  38.70  &   112.1  &  31.3  &  0.22  &  A-SM1  \\  
A-15  &    0.4  &    8.9  &  3.89  &  0.35  &  27.00  &  27.30  &    25.1  &  27.1  &  0.13  &  VLA~1623  \\  
\hline 
B1-1  &   38.9  &  -33.0  &  4.00  &  0.73  &  31.80  &  32.20  &    50.6  &  32.0  &  0.37  &  B1-MM4  \\  
B1-4  &   -0.8  &    0.0  &  3.75  &  0.60  &  27.00  &  27.40  &    86.3  &  27.2  &  0.25  &  B1-MM3  \\  
B1-20  &  -22.3  &   53.8  &  4.24  &  0.45  &  27.00  &  58.10  &    41.8  &  34.6  &  0.15  &  B1-MM2  \\  
\hline 
B2-1  &   67.8  &    9.3  &  4.43  &  1.09  &  30.80  &  48.20  &   104.2  &  36.7  &  0.29  &  B2-MM15  \\  
B2-2  &   56.5  &   50.5  &  3.91  &  0.92  &  46.40  &  27.30  &    -6.3  &  33.2  &  0.35  &  B2-MM13  \\  
B2-3  &   16.6  &    1.0  &  4.17  &  0.78  &  34.40  &  50.20  &   109.2  &  40.1  &  0.31  &  $\sim$ B2-MM8  \\  
B2-4  &   72.3  &  -39.3  &  4.02  &  0.77  &  30.20  &  43.20  &   134.7  &  35.0  &  0.36  &    \\  
B2-5  &  -30.0  &  -10.2  &  3.76  &  0.66  &  51.70  &  28.60  &   -36.5  &  35.4  &  0.37  &  B2-MM5  \\  
B2-6  &   49.4  &  -33.3  &  3.53  &  0.54  &  27.10  &  35.80  &   121.9  &  30.6  &  0.46  &    \\  
B2-7  &  -20.9  &   19.9  &  3.41  &  0.49  &  27.00  &  46.40  &    94.9  &  33.0  &  0.24  &    \\  
B2-8  &   83.7  &   45.4  &  4.16  &  0.51  &  27.00  &  33.90  &    17.7  &  29.9  &  0.28  &  $\sim$ B2-MM16,17  \\  
B2-10  &  -70.5  &  -49.9  &  3.79  &  0.45  &  27.00  &  48.40  &   171.4  &  33.4  &  0.22  &  B2-MM3  \\  
B2-11  &   10.1  &  -36.9  &  4.41  &  0.45  &  27.00  &  27.30  &    78.7  &  27.1  &  0.23  &  $\sim$ B2-MM7,10  \\  
\hline 
C-1  &   -2.4  &   -6.2  &  3.67  &  0.93  &  63.70  &  32.10  &    15.1  &  40.5  &  0.32  &  C-MM6  \\  
C-2  &  -39.1  &   21.4  &  3.92  &  0.76  &  50.50  &  39.80  &    11.7  &  44.2  &  0.31  &  C-MM3, C-MM5  \\  
C-3  &  -21.5  &   72.0  &  3.78  &  0.42  &  27.00  &  60.70  &   140.6  &  34.9  &  0.19  &    \\  
\hline 
E-1  &    7.7  &    3.3  &  4.52  &  0.61  &  50.30  &  34.00  &     5.5  &  39.8  &  0.25  &  E-MM2d  \\  
E-4  &   80.6  &  -10.9  &  4.28  &  0.37  &  27.00  &  27.30  &    24.7  &  27.2  &  0.28  &  E-MM4  \\  
\hline 
F-1  &   33.7  &  -22.5  &  4.15  &  0.82  &  66.10  &  30.60  &     6.0  &  39.3  &  0.24  &  $\sim$ YLW~15  \\  
F-2  &   -8.1  &   29.9  &  4.54  &  0.51  &  27.00  &  42.10  &    60.3  &  32.1  &  0.26  &  $\sim$ F-MM1  \\  
F-4  &    5.2  &  -12.9  &  4.66  &  0.41  &  27.00  &  35.30  &    62.0  &  30.3  &  0.31  &  $\sim$ F-MM2  \\  
\hline 
\end{tabular} 
\begin{list}{}{}
\item[]{Notes: In col~[1], the first part of the object name indicates the DCO$^+$ core in which the object is embedded.
Columns [2] and [3] are the offsets of the object with respect to the core position given in the caption of Fig.~\ref{fig_oph_mapn2h+}.
Columns [4] and [5] are the center velocity and peak intensity, respectively.
Columns [6]  and [7] give the major and minor angular diameters of the two-dimensional Gaussian fitted to 
the object, and col.~[8] is the position angle of the minor axis. 
Columns [9] and [10] are the typical FWHM size and FWHM line width, respectively.} 
\item[$(a)$]{N$_2$H$^+$ peak N6 discussed by \citet{DiFrancesco04}.} 
\end{list} 
\end{table*}

In addition, we also used the {\it Gaussclumps} fitting procedure of \citet{Stutzki90} 
\citep[see also][]{Kramer98} to decompose each of our background-subtracted OTF N$_2$H$^+$(101-012), H$^{13}$CO$^+$(1-0), 
and DCO$^+$(2-1) data cubes into a series of Gaussian-shaped ``clumps''. The significant ($> 5\, \sigma$) clumps identified with 
{\it Gaussclumps} in the various DCO$^+$ cores of L1688 are shown as ellipses in 
Fig.~\ref{fig_oph_mapn2h+} and Fig.~\ref{fig_oph_maph13co+}. 
Their main characteristics are given in Table~\ref{t:gaussclumps_results}. 
In this way, a total of 17 starless 1.2~mm condensations and 3 Class~0/Class~I protostars (VLA~1623, LFAM~26, GY~210)  
were found to have well-defined N$_2$H$^+$, H$^{13}$CO$^+$, or DCO$^+$ counterparts in position-velocity ($l-v$) space.
 The relevance of such a {\it Gaussclumps} decomposition is that it further helped us discriminate between the line emission 
arising from the compact condensations themselves (with well-defined ``positions'' in $l-v$ space) and the line emission from 
the surrounding, more extended dense gas (with less-well-defined ``positions'' in $l-v$ space). 
In particular, for six of the above-mentioned eight condensations with double N$_2$H$^+$ HFS components, the {\it Gaussclumps} decomposition 
allowed us to reject one of the two components and to identify the component most likely associated with the compact 1.2~mm continuum 
object. 
 In Oph~B1, for instance, two HFS components were detected toward B1-MM2, B1-MM3, and B1-MM4 prior to background 
subtraction (cf. Table~\ref{t:fithfs_results}), but only one of these two components was found to be associated with each condensation 
after running {\it Gaussclumps} on the background-subtracted N$_2$H$^+$(101-012) data cubes (cf. Table~\ref{t:gaussclumps_results}).
In the following, we will consider the multiple components detected toward the remaining two 
condensations with double HFS components (B2-MM2 and B2-MM4) 
as representative of independent objects when we discuss the statistics of linewidths (Sect. ~\ref{obs_virial})
and relative motions (Sect. ~\ref{sec_oph_vlsr}) between condensations.

\subsection{Linewidths and virial masses of the protocluster condensations}
\label{obs_virial}

\begin{table*} 
\caption{Line-of-sight velocity dispersions and virial mass estimates for the 41 starless 1.2~mm condensations positively detected in N$_2$H$^+$.} 
\label{t:veldisp_results} 
\vspace*{0.5ex} 
\begin{tabular}{lcccccccccccc} 
\hline\hline 
 & & & & & \multicolumn{4}{c}{Prior to background subtraction} &  \multicolumn{4}{c}{After background subtraction}  \\  
Source & FWHM & $T_d$$^b$ & $M_{1.2mm}$$^b$ & $\sigma_T(\mu)$ & $\sigma_{NT}$ & $\frac{\sigma_{NT}}{\sigma_T}$ & $M_{vir}$ & $\frac{M_{vir}}{M_{1.2mm}}$ & $\sigma_{NT}$ & $\frac{\sigma_{NT}}{\sigma_T}$ & $M_{vir}$ & $\frac{M_{vir}}{M_{1.2mm}}$ \\  
 & \scriptsize{(AU $\times$ AU)} & \scriptsize{(K)} & \scriptsize{(M$_\odot$)} & \scriptsize{(km~s$^{-1}$)} & \scriptsize{(km~s$^{-1}$)} & & \scriptsize{(M$_\odot$)} & & \scriptsize{(km~s$^{-1}$)} &  & \scriptsize{(M$_\odot$)} & \\  
(1) & (2) & (3) & (4) & (5) & (6) & (7) & (8) & (9) & (10) & (11) & (12) & (13) \\  
\hline 
A3-MM1  & $<$        400  &         12  &     0.21  &    0.21  &     0.37  &      1.8  & $<$    0.24  &  $<$     1.2  &  & & & \\  
A-MM4  &       4000  $\times$        1400  &         12  &     0.29  &    0.21  &     0.18  &      0.9  &    0.60  &      2.1  &    0.15  &      0.7  &    0.52  &      1.8  \\  
A-MM5  &       3700  $\times$        2900  &         12  &     0.48  &    0.21  &     0.15  &      0.7  &    0.72  &      1.5  &    0.05  &      0.2  &    0.50  &      1.1  \\  
SM1N  &       3000  $\times$        1800  &         20  &     1.30  &    0.27  &     0.19  &      0.7  &    0.85  &      0.7  &    0.17  &      0.6  &    0.78  &      0.6  \\  
SM1  &       6100  $\times$        2100  &         20  &     3.20  &    0.27  &     0.24  &      0.9  &    1.57  &      0.5  &    0.26  &      1.0  &    1.68  &      0.5  \\  
A-MM6  &       3200  $\times$        2700  &         20  &     0.40  &    0.27  &     0.28  &      1.1  &    1.51  &      3.8  &  & & & \\  
SM2  &       6200  $\times$        3400  &         20  &     1.30  &    0.27  &     0.19  &      0.7  &    1.67  &      1.3  &    0.15  &      0.5  &    1.43  &      1.1  \\  
A-MM8  &       2900  $\times$        2100  &         20  &     0.13  &    0.27  &     0.14  &      0.5  &    0.77  &      5.9  &    0.12  &      0.4  &    0.71  &      5.5  \\  
A-S  & $<$        400  &         12  &     0.17  &    0.21  &     0.08  &      0.4  & $<$    0.07  &  $<$     0.4  &  & & & \\  
\hline                                                                 
B1-MM1  & $<$        400  &         12  &     0.10  &    0.21  &     0.15  &      0.7  & $<$    0.09  &  $<$     0.9  &  & & & \\  
B1-MM2$^a$ &       3000  $\times$        2100  &         12  &     0.17  &    0.21  &     0.11  &      0.5  &    0.46  &      2.7  &    0.07  &      0.3  &    0.40  &      2.4  \\  
B1-MM3$^a$ &       1800  $\times$        1300  &         12  &     0.16  &    0.21  &     0.13  &      0.6  &    0.31  &      1.9  &    0.05  &      0.3  &    0.23  &      1.5  \\  
B1-MM4$^a$ &       4600  $\times$        3200  &         12  &     0.21  &    0.21  &     0.13  &      0.6  &    0.78  &      3.7  &    0.06  &      0.3  &    0.61  &      2.9  \\  
\hline                                                                 
B1B2-MM1  &       2700  $\times$        1800  &         12  &     0.10  &    0.21  &     0.15  &      0.7  &    0.49  &      4.9  &    0.07  &      0.3  &    0.35  &      3.5  \\  
\hline                                                                 
B2-MM1  & $<$        400  &         12  &     0.14  &    0.21  &     0.15  &      0.7  & $<$    0.09  &  $<$     0.6  &    0.05  &      0.3  & $<$    0.06  &  $<$     0.4  \\  
B2-MM2  &       4500  $\times$        2400  &         12  &     0.47  &    0.21  &     0.32  &      1.5  &    1.58  &      3.4  &  & & & \\  
 &   &   &   &    0.21  &     0.08  &      0.4  &    0.55  &      1.2  &  & & & \\  
B2-MM3$^a$ & $<$        400  &         12  &     0.12  &    0.21  &     0.22  &      1.1  & $<$    0.12  &  $<$     1.0  &  & & & \\  
B2-MM4  &       2100  $\times$         960  &         12  &     0.27  &    0.21  &     0.16  &      0.8  &    0.32  &      1.2  &    0.09  &      0.5  &    0.25  &      0.9  \\  
 &   &   &   &    0.21  &     0.13  &      0.7  &    0.29  &      1.1  &  & & & \\  
B2-MM5$^a$ &       2200  $\times$         960  &         12  &     0.26  &    0.21  &     0.20  &      1.0  &    0.41  &      1.6  &    0.12  &      0.6  &    0.27  &      1.1  \\  
B2-MM6  &       4300  $\times$        2700  &         12  &     0.78  &    0.21  &     0.29  &      1.4  &    1.47  &      1.9  &    0.17  &      0.8  &    0.82  &      1.1  \\  
B2-MM7  & $<$        400  &         12  &     0.23  &    0.21  &     0.33  &      1.6  & $<$    0.20  &  $<$     0.9  &  & & & \\  
B2-MM8  &       4000  $\times$        4000  &         12  &     1.50  &    0.21  &     0.24  &      1.2  &    1.36  &      0.9  &    0.17  &      0.8  &    0.96  &      0.6  \\  
B2-MM9  &       1600  $\times$         960  &         12  &     0.31  &    0.21  &     0.27  &      1.3  &    0.48  &      1.6  &  & & & \\  
B2-MM10  &       3400  $\times$        2200  &         12  &     0.60  &    0.21  &     0.23  &      1.1  &    0.90  &      1.5  &  & & & \\  
B2-MM11  & $<$        400  &         12  &     0.15  &    0.21  &     0.43  &      2.1  & $<$    0.31  &  $<$     2.0  &    0.26  &      1.3  & $<$    0.15  &  $<$     1.0  \\  
B2-MM12  &       2100  $\times$        1300  &         12  &     0.39  &    0.21  &     0.23  &      1.1  &    0.52  &      1.3  &    0.08  &      0.4  &    0.28  &      0.7  \\  
B2-MM13  & $<$        400  &         12  &     0.19  &    0.21  &     0.26  &      1.3  & $<$    0.15  &  $<$     0.8  &  & & & \\  
B2-MM14  &       2100  $\times$        1800  &         12  &     0.43  &    0.21  &     0.32  &      1.6  &    0.96  &      2.2  &  & & & \\  
B2-MM15  & $<$        400  &         12  &     0.17  &    0.21  &     0.14  &      0.7  & $<$    0.08  &  $<$     0.5  &    0.07  &      0.4  & $<$    0.06  &  $<$     0.4  \\  
B2-MM16  &       2700  $\times$        1300  &         12  &     0.35  &    0.21  &     0.26  &      1.3  &    0.70  &      2.0  &    0.17  &      0.8  &    0.45  &      1.3  \\  
B2-MM17  & $<$        400  &         12  &     0.23  &    0.21  &     0.26  &      1.2  & $<$    0.15  &  $<$     0.6  &    0.12  &      0.6  & $<$    0.08  &  $<$     0.3  \\  
\hline                                                                 
C-MM2  & $<$        400  &         12  &     0.12  &    0.21  &     0.17  &      0.8  & $<$    0.10  &  $<$     0.8  &  & & & \\  
C-MM3  &       5400  $\times$         640  &         12  &     0.23  &    0.21  &     0.12  &      0.6  &    0.35  &      1.5  &  & & & \\  
C-MM4  &       2400  $\times$        1400  &         12  &     0.16  &    0.21  &     0.13  &      0.6  &    0.37  &      2.3  &  & & & \\  
C-MM5  & $<$        400  &         12  &     0.10  &    0.21  &     0.12  &      0.6  & $<$    0.08  &  $<$     0.8  &    0.08  &      0.4  & $<$    0.07  &  $<$     0.7  \\  
C-MM6  &       4000  $\times$        3700  &         12  &     0.33  &    0.21  &     0.13  &      0.6  &    0.76  &      2.3  &    0.09  &      0.4  &    0.65  &      2.0  \\  
C-MM7  & $<$        400  &         12  &     0.13  &    0.21  &     0.11  &      0.5  & $<$    0.07  &  $<$     0.6  &  & & & \\  
\hline                                                                 
E-MM2d  &       4200  $\times$        2700  &         12  &     0.63  &    0.21  &     0.11  &      0.5  &    0.61  &      1.0  &    0.09  &      0.4  &    0.58  &      0.9  \\  
E-MM4  &       6900  $\times$        5300  &         12  &     0.61  &    0.21  &     0.13  &      0.6  &    1.20  &      2.0  &    0.09  &      0.4  &    1.04  &      1.7  \\  
\hline                                                                 
F-MM1  &       4800  $\times$        2600  &         12  &     0.35  &    0.21  &     0.19  &      0.9  &    0.95  &      2.7  &    0.13  &      0.6  &    0.70  &      2.0  \\  
F-MM2$^a$ &       2700  $\times$        1600  &         12  &     0.17  &    0.21  &     0.14  &      0.7  &    0.43  &      2.6  &    0.12  &      0.6  &    0.39  &      2.3  \\  
\hline 
\end{tabular} 
\begin{list}{}{} 
 \item[$(a)$]{Condensation with two velocity components in N$_2$H\/$^+$ (see Table~\ref{t:fithfs_results}). Only the velocity component identified with \textit{Gaussclumps} (see Table~\ref{t:gaussclumps_results}) is considered here.} 
 \item[$(b)$]{For some condensations, the dust temperatures adopted here to estimate $M_{1.2mm}$ from the measured 1.2mm flux differ slightly from those assumed by 
\citeauthor*{Motte98}. For simplicity, we used $T_d = 12$~K for all condensations except SM1N, SM1, A-MM6, SM2, A-MM8 in Oph~A, which are likely warmer 
\citep[$T_d = 20$~K -- cf.][]{Andre93}. 
%$T_d = 12$~K is consistent with the typical mass-averaged temperature found by dust radiative transfer models of starless cores \citep[cf.][]{Andre03}. 
The mass spectra shown in Fig.~\ref{fig_preimf} and Fig.~\ref{fig_wcmd} were obtained using the dust temperatures quoted here.} 
\end{list} 
\end{table*}

Based on the measured N$_2$H$^+$(1-0) linewidths (Table~\ref{t:fithfs_results}),
we can calculate the nonthermal component of the 
line-of-sight velocity dispersion \citep[cf.][]{Myers99}
and estimate a virial mass for each condensation. 
Table~\ref{t:veldisp_results} gives the deconvolved (FWHM) diameter 
(col.~[2], \citeauthor*{Motte98})
and gas~$+$~dust mass derived from the 1.2~mm continuum (col.~[4],
\citeauthor*{Motte98}), the assumed gas/dust temperature (col.~[3], cf. 
\citeauthor*{Motte98}), the thermal velocity dispersion $\sigma_T (\mu) $ 
for a particle of mean molecular weight $\mu  = 2.33$ (col.~[5]), 
the one-dimensional nonthermal velocity dispersion  $\sigma_{NT}$ 
obtained from the N$_2$H$^+$(1-0) linewidth (col.~[6]), the nonthermal to 
thermal velocity dispersion ratio $\sigma_{NT}/\sigma_{T}$ (col.~[7]),
the estimated virial mass $M_{vir}$ (col.~[8]), and the virial mass ratio 
$\alpha_{vir} \equiv M_{vir}/M_{1.2mm}$ (col.~[9]) for each observed object.
Columns [10] to [13] list the values of the same parameters when estimated from the 
background-subtracted spectra. 
 For some condensations, the dust temperatures (col.~[3]) adopted  to estimate $M_{1.2mm}$ 
(col.~[4]) from the measured 1.2mm flux differ slightly from those assumed by 
\citeauthor*{Motte98}. Here, we used $T_d = 12$~K for all condensations except SM1N, SM1, 
A-MM6, SM2, A-MM8 in Oph~A, which are likely warmer ($T_d = 20$~K) 
due to their proximity to the B3 star S1 \citep[cf.][]{Andre93}.  Our adopted default 
$T_d = 12$~K is consistent with the typical mass-averaged temperature found by 
dust radiative transfer models of starless cores \citep[cf.][]{Andre03,Stamatellos03}. 
In agreement with the radial profiles measured by \citeauthor*{Motte98}, the values quoted for $M_{vir}$ and $\alpha_{vir}$ 
assume that the density structure of the sources approaches that of centrally-condensed
spheres with outer profiles such as $\rho \propto r^{-2}$, for which $M_{vir} \approx 3\, r_{cond}\, \frac{\sigma_{tot} ^2}{G}$. 
Here, $\sigma_{tot} = \sqrt{\sigma_T(\mu)^2 + \sigma_{NT}^2}$ is the total 
(thermal $+$ nonthermal) line-of-sight velocity dispersion and $r_{cond}$ is the condensation outer radius,
which we take to be twice the geometrical mean of the deconvolved major and minor HWHM radii 
measured in the 1.2~mm dust continuum 
(i.e., $\sqrt{\rm{FWHM_a}Ê\times \rm{FWHM_b}}$ -- cf. col.~[2] of Table~\ref{t:veldisp_results} and Table~2 of \citeauthor*{Motte98}).
For unresolved condensations, the quoted $M_{vir}$ and $\alpha_{vir}$ values are only upper limits 
obtained under the assumption that the deconvolved FWHM diameter is 400~AU (i.e., $\sim$~one fourth of  
the HPBW spatial resolution of the 1.2~mm continuum observations). 
Note also that our N$_2$H$^+$(1-0) observations provide the line-of-sight velocity dispersion 
averaged over a $\sim  26$\arcsec ~beam, while the FWHM angular size of the dust condensations 
ranges from $< 15\arcsec $ to $\sim 30 \arcsec $.
However, since both $M_{vir}$ and $M_{1.2mm}$ are derived for a diameter $\sim$ twice the FWHM size,
our method of estimating $\alpha_{vir}$ should be reasonably accurate.
Some of the secondary components listed in 
Table~\ref{t:fithfs_results} were ignored in Table~\ref{t:veldisp_results}, 
based on the \textit{Gaussclumps} identifications summarized in 
Table~\ref{t:gaussclumps_results}.

The measured N$_2$H\/$^+$(1-0) linewidths are narrow, indicative of 
small internal velocity dispersions. 
Prior to background subtraction, 
the average one-dimensional nonthermal 
velocity dispersion, $\sigma_{NT}$, is estimated to be 
$\sim$ 0.20 $\pm$ 0.08~km~s$^{-1}$
among the  43 components listed in Table~\ref{t:veldisp_results}.
 More precisely, the average one-dimensional nonthermal 
velocity dispersion of the condensations ranges from 0.12 $\pm$ 0.02~km~s$^{-1}$ in Oph~E,
0.13 $\pm$ 0.02~km~s$^{-1}$ in Oph~C,  
0.15 $\pm$ 0.04~km~s$^{-1}$ in Oph~B1/B1B2, 
0.17 $\pm$ 0.04~km~s$^{-1}$ in Oph~F, 
to 0.20 $\pm$ 0.08~km~s$^{-1}$ in Oph~A,  and 
0.24 $\pm$ 0.08~km~s$^{-1}$ in Oph~B2. 
For comparison, the isothermal sound speed or one-dimensional thermal velocity 
dispersion is $\sigma_{T} \sim $~0.21--0.27~km~s$^{-1}$ for gas temperatures 
of 12--20~K. Therefore, the condensations of Oph~A, Oph~B1, Oph~C, Oph~E, and 
Oph~F are characterized by only {\it subsonic} 
levels of internal turbulence with 
$\sigma_{NT}/\sigma_{T} \sim $~0.6--0.8 on average, while  
the condensations of Oph~B2 have at most ``transonic'' internal turbulence 
with $\sigma_{NT}/\sigma_{T} \sim 1.1  < 2$ on average. 
This is especially true since the measured nonthermal component of the N$_2$H\/$^+$(1-0) 
linewidth may overestimate the intrinsic level of internal turbulence in the presence of 
infall motions and/or rotation (see Sect.~\ref{sub_oph_infall} and  
\ref{sub_oph_gradv}).

In contrast, the line-of-sight velocity dispersions measured  
on the scale of the parent DCO$^+$ cores are significantly larger 
\citep*[cf. Table~2 of][]{Loren90}, with corresponding values of $\sigma_{NT}$ ranging 
from $\sim 0.17$ km~s$^{-1}$ in Oph~E, 0.2--0.3 km~s$^{-1}$ in Oph~C and
Oph~F, to 0.35--0.5~km~s$^{-1}$ in Oph~A, Oph~B1, and Oph~B2. 
Furthermore, the values of $\sigma_{NT}$ 
derived from NH$_3$ observations of the cores tend to be larger, by up to a factor of 2, 
than those inferred from DCO$^+$ \citep*[][; A. Wootten, private communication]{Loren90}. For instance, 
$\sigma_{NT} = 0.5 \pm 0.1$ km~s$^{-1}$ is estimated in Oph~B1 from NH$_3$ 
\citep[][]{Zeng84}, compared to $\sigma_{NT} = 0.4$ km~s$^{-1}$ from DCO$^+$.
Thus, the internal
velocity dispersions derived here for the compact condensations 
are typically a factor of $\sim 2$ smaller than the nonthermal velocity dispersions 
of their parent dense cores.
We conclude that the level of turbulence decreases from marginally supersonic on the
scale of the DCO$^+$ cores to transonic or marginally subsonic on the scale 
of the prestellar condensations. 

The above results are reminiscent of the transition to ``coherence'' observed by \citet{Goodman98} in 
isolated starless cores. They suggest that, at least in nearby cluster-forming clouds, 
the initial conditions for individual protostellar collapse are ``coherent'' and largely free of turbulence. 
They are also consistent with scenarios, such as the ``kernel'' model of \citet{Myers98}, 
according to which protocluster condensations form by dissipation of MHD turbulence on small 
scales within massive, turbulent cores \citep[see also][]{Nakano98}.

%%%
%
%\section{Systematic motions within the DCO$^+$ cores of L1688}
\section{Motions within the DCO$^+$ cores of L1688}
\label{oph_kin}

\subsection{Evidence of gravitational infall motions}
\label{sub_oph_infall}

\begin{figure} [!ht]
\vspace*{0.0 cm}
\centerline{\resizebox{1.05\hsize}{!}{\includegraphics[angle=0]{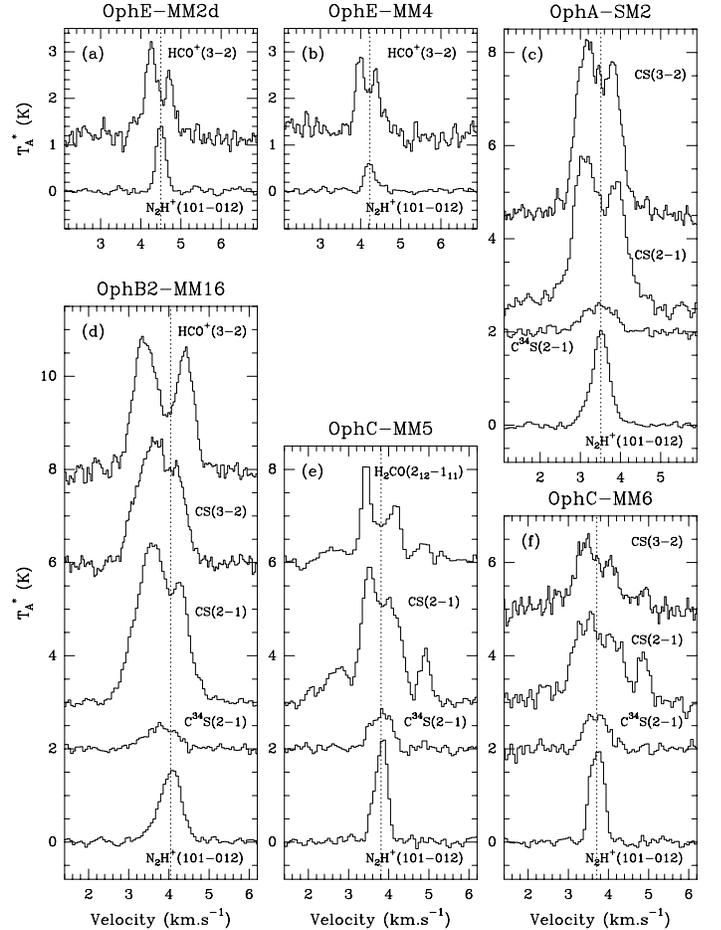}}}
\caption[Infall profiles in 6 $\rho$~Oph condensations]{Spectroscopic signatures of 
infall motions observed toward 6  Ophiuchus starless condensations: (a) OphE-MM2d, (b) OphE-MM4, 
(c) OphA-SM2, (d) OphB2-MM16, (e) OphC-MM5, and (f) OphC-MM6.  
On each panel, the vertical dotted line marks the source systemic velocity as derived from a 
Gaussian HFS fit to the observed N$_2$H\/$^+$(1-0) spectrum.}
\label{fig_oph_infall}
\end{figure}

Our data show the classical  spectroscopic signature of infall motions 
\citep*[cf.][]{Evans99,Myers00b} toward at least six 1.2~mm continuum condensations, 
where optically thick lines 
such as CS(2--1), CS(3--2), H$_2$CO($2_{12} - 1_{11}$), 
and/or HCO$^+$(3--2) are double-peaked 
with a stronger blue peak, while low-optical-depth lines such as N$_2$H$^+$(101--012) 
and C$^{34}$S(2--1) 
peak in the dip between the blue and red peaks of the optically thick tracers 
(cf. Fig.~\ref{fig_oph_infall})\footnote{The optically thick spectra observed toward C-MM5 and C-MM6 
show two fainter, additional peaks at velocities $\sim 2.8$ and 
$\sim4.9$ km~s$^{-1}$. These additional velocity components are also 
observed in C$^{18}$O(2-1) but are not seen in N$_2$H$^+$(1-0). 
They likely arise from the ambient cloud rather than from 
C-MM5 et C-MM6 themselves.}. 
Such asymmetric line profiles skewed to the blue in optically thick tracers 
are produced when a gradient in excitation temperature toward source center 
is combined with inward motions \citep[e.g.][]{Leung77}.
Here, among a total of 25 starless condensations observed in at least one optically thick
infall tracer and one optically thin transition, blue infall profiles are clearly 
observed in 6 condensations (A-SM2, B2-MM16, C-MM5, C-MM6, E-MM2d, and E-MM4, 
see Fig.~\ref{fig_oph_infall}), 
tentatively observed in 10 other condensations 
(A-MM4, A-MM5, A-SM1, A-MM8, C-N, B2-MM12, B2-MM13, B2-MM14, B2-MM15, B2-MM17), 
and not seen in the remaining 9 sources 
(A-MM6, A-SM1N, F-MM1, F-MM2, B1-MM4, B2-MM2, B2-MM6, B2-MM8, B2-MM10).

In the case of  %E-MM2d, 
A-SM2 and C-MM6, 
the absorption dip observed in the optically thick line profiles is slightly 
redshifted with respect to the LSR velocity traced by the optically 
thin lines, 
by %$\sim 0.05$, 
$\sim 0.1$ and $\sim$~0.1-0.2 km~s$^{-1}$, respectively.
This suggests that the outer gas layers responsible for the absorption dip in these 
two condensations are characterized by significant inward velocities $\sim 0.1$~km~s$^{-1}$ 
\citep[see, e.g., discussion in Sect. ~3.3 of][]{Belloche02}.
By contrast, the absorption dip is not redshifted (nor blueshifted) toward  
E-MM2d, E-MM4, B2-MM16, and C-MM5, suggesting that only comparatively 
deeper layers undergo infall motions in the latter sources.

%\subsection{Large-scale velocity gradients}
\subsection{Large-scale velocity structure}
\label{sub_oph_gradv}

\begin{figure*} [!ht]
\centerline{\resizebox{1.0\hsize}{!}{\includegraphics[angle=270]{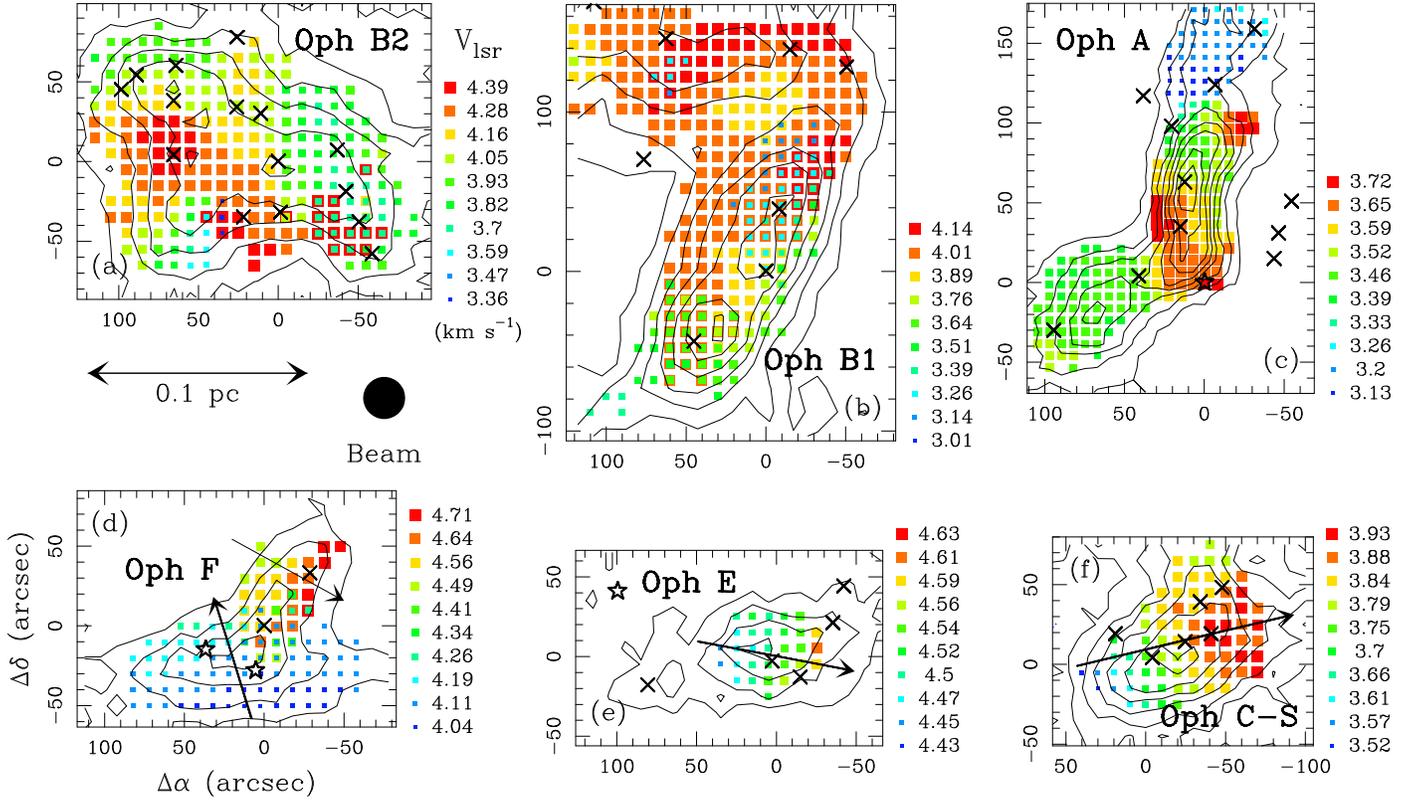}}}
\caption{ 
Centroid velocity maps of Oph~B2 (a), Oph~B1 (b), Oph~A (c), Oph~F (d), 
Oph~E (e), and Oph~C-S (f) as derived from HFS fits with one or two velocity 
components to the N$_2$H\/$^+$(1-0) spectra with a signal-to-noise ratio 
larger than $\sim$ 5 (filled squares of varying sizes and colors). 
The (0,0) positions are given in the caption of Fig.~\ref{fig_oph_mapn2h+}.
The linear size of the squares increases, and their color changes from blue to red, as 
$V_{lsr}$ increases. The underlying contours represent the same 
N$_2$H\/$^+$(1-0) integrated intensity maps as shown in 
Fig.~\ref{fig_oph_mapn2h+}a--f. The direction of the fitted velocity gradients
in Oph~F, Oph~C-S, and Oph~E are shown by arrows in panels (d), (e), and (f) 
(see Table~\ref{t:ophcef_gradv}). In Oph~F, two velocity components were
fitted separately and the arrows show the two corresponding velocity gradients 
associated with components Oph~F1 at $v \sim 4.2$ km~s$^{-1}$ and 
Oph~F2 at $v \sim 4.6$ km~s$^{-1}$, respectively.
Crosses mark the 1.2~mm continuum positions of 
starless condensations (\citeauthor*{Motte98}), 
while stars mark the positions of protostars (VLA~1623, YLW~15/IRS~43, 
CRBR~85, and WL~19).}
\label{fig_oph_velcore}
\end{figure*}

In order to investigate the presence of systematic velocity gradients, such as rotational gradients,
within the DCO$^+$ cores  of L1688, we plot maps of the N$_2$H$^+$(1-0) centroid velocity across  
Oph~A, Oph~B1, Oph~B2, Oph~C-S, Oph~E, and Oph~F in 
Fig.~\ref{fig_oph_velcore}a--f. 
 These centroid velocity maps were derived from the original N$_2$H$^+$(1-0) data cubes 
before background subtraction.  
No clear, large-scale velocity gradient can be seen in the velocity maps of 
Oph~A, Oph~B1, and Oph~B2,  and the underlying kinematic structure cannot be described as 
a simple superposition of a small number of velocity gradients. 
In Oph~A, which resembles a curved filament in the integrated intensity map, the two ends
of the filament are characterized by lower systemic velocities ($V_{lsr} \sim 3.2$ and $\sim 3.5$ km~s$^{-1}$)
than the central region ($V_{lsr} \sim 3.7$ km~s$^{-1}$), dominated by the objects A-SM1 and VLA~1623. 
In Oph~B2, the two condensations B2-MM10 and B2-MM15 stand out with significantly higher 
systemic velocities ($V_{lsr} \sim 4.4$ km~s$^{-1}$) than the bulk of the 
core ($V_{lsr} \sim 3.7-4.1$ km~s$^{-1}$). 
In Oph~B1, two velocity components are detected on the lines of sight
to B1-MM2, B1-MM3, and B1-MM4. Based on the results of our 
\textit{Gaussclumps} analysis (see Table~\ref{t:gaussclumps_results}), we believe that 
only the $\sim 4.0$ km~s$^{-1}$ component, connected to Oph~B2, 
is physically associated with the compact condensations B1-MM2, B1-MM3, and B1-MM4.

\begin{table*} [!t]
 \caption{Results of velocity gradient fitting}
 \label{t:ophcef_gradv}
 \vspace*{2.ex}
 \centering
 \footnotesize
 \begin{tabular}{llcccccc}
 \hline
 \hline
 \multicolumn{1}{c}{Source} & \multicolumn{1}{c}{Transition} & \multicolumn{1}{c}{Size$^a$} & \multicolumn{1}{c}{$n_{ind}^b$} & \multicolumn{1}{c}{$V_0^c$} & \multicolumn{1}{c}{$\|\vec{\nabla} V\|^c$} & \multicolumn{1}{c}{$P.A.^d$} & \multicolumn{1}{c}{rms$^e$} \\
 \multicolumn{1}{c}{} & \multicolumn{1}{c}{} & \multicolumn{1}{c}{\tiny ($\arcsec \times \arcsec$)} & \multicolumn{1}{c}{} & \multicolumn{1}{c}{\tiny (km~s$^{-1}$)} & \multicolumn{1}{c}{\tiny (km~s$^{-1}$pc$^{-1}$)} & \multicolumn{1}{c}{($^\circ$)} & \multicolumn{1}{c}{\tiny (km~s$^{-1}$)} \\*[0.4ex]
(1) & (2) & (3) & (4) & (5) & (6) & (7) & (8) \\
 \hline
 Oph~C-S & N$_2$H\/$^+$(1-0)      & $120 \times 70$ & 14 & 3.72 & 4.4 &  -77 & 0.042 \\
 \hline
 Oph~F1 & N$_2$H\/$^+$(1-0)       & $150 \times 50$ & 16 & 4.17 & 3.8 &   17 & 0.026 \\
 Oph~F2 & N$_2$H\/$^+$(1-0)       & $50 \times 80$  &  5 & 4.60 & 6.3 & -119 & 0.028 \\
 \hline
 Oph~E  & H\/$^{13}$CO\/$^+$(1-0) & $180 \times 80$ & 19 & 4.47 & 5.0 & -106 & 0.052 \\
        & DCO\/$^+$(2-1)          & $170 \times 70$ & 43 & 4.53 & 5.1 & -104 & 0.051 \\
        & DCO\/$^+$(3-2)          & $160 \times 60$ & 68 & 4.48 & 5.6 & -113 & 0.044 \\
        & Average                 &                 & & & 5.2(3) & -108(4) & \\
% E-MM2d & H\/$^{13}$CO\/$^+$(1-0) & $80 \times 80$  &  8 & 4.48 & 5.0 & -113 & 0.034 \\
%        & DCO\/$^+$(2-1)          & $70 \times 70$  & 17 & 4.52 & 5.0 & -122 & 0.030 \\
%        & DCO\/$^+$(3-2)          & $50 \times 50$  & 18 & 4.46 & 5.6 & -143 & 0.036 \\
%        & \textbf{average}           &                 & & & \textbf{5.2$\pm$0.3} & \textbf{-126$\pm$15} & \\*[0.9ex]
% E-MM4  & H\/$^{13}$CO\/$^+$(1-0) & $60 \times 50$  &  5 & 4.56 & 6.7 &  -57 & 0.029 \\
%        & DCO\/$^+$(2-1)          & $50 \times 50$  & 10 & 4.63 & 7.4 &  -47 & 0.024 \\
%        & DCO\/$^+$(3-2)          & $40 \times 30$  & 12 & 4.56 & 7.1 &  -41 & 0.019 \\
%        & \textbf{average}           &                 & & & \textbf{7.1$\pm$0.4} & \textbf{-48$\pm$8} & \\*[0.9ex]
 \hline
 \end{tabular}
 \vspace*{0.5ex}
 \begin{list}{}{}
  \item[]{Notes: the numbers in parentheses indicate the uncertainty in units 
    of the last digit.}
  \item[$(a)$]{Approximate diameter of the region over which a velocity 
    gradient was fitted}
  \item[$(b)$]{Number of fully independent points used in the gradient 
    fitting.}
  \item[$(c)$]{The centroid velocity maps were fitted with the function 
    $V_0 + \vec{\nabla} V.\vec{\Delta X}$, with $\vec{\Delta X}$ the position 
    vector measured from the (0,0) position.}
  \item[$(d)$]{Position angle of the direction of the velocity gradient 
    $\vec{\nabla} V$.}
  \item[$(e)$]{rms residual of the fit.}
%C'est le rapport de la surface de la r\'egion s\'electionn\'ee sur la surface du lobe $\pi \times FWHM^2/4$.}
 \end{list}
 \normalsize
\end{table*}

Well-defined N$_2$H$^+$(1-0) velocity gradients were identified toward Oph~C-S, Oph~E, and Oph~F, 
and their amplitudes estimated using the least-squares fitting method of \citet{Goodman93}.
%(see discussion in \S ~\ref{sub_ophcs_vel}  and  \S ~\ref{sub_ophe_vel} below).
In particular, Oph~C-S is characterized by a systematic projected velocity gradient of mean 
value $\|\vec{\nabla} V\| \sim 4.4$ km~s$^{-1}$~pc$^{-1}$, 
with velocities increasing along the direction of position angle $P.A. \sim -77^\circ$ in 
the plane of the sky (see Table~\ref{t:ophcef_gradv} and Fig.~\ref{fig_oph_velcore}f).

In Oph~E, our N$_2$H$^+$(1-0) data suggest the presence of a velocity gradient 
$\|\vec{\nabla} V\| \sim 3.0$ km~s$^{-1}$~pc$^{-1}$ at $P.A. \sim -101^\circ$ 
(cf. Fig.~\ref{fig_oph_velcore}e). The signal-to-noise ratio in our N$_2$H$^+$(1-0) map is however 
insufficient to constrain the nature of the velocity field. 
Fortunately, we can also use our higher signal-to-noise maps taken in H$^{13}$CO$^+$(1-0), 
DCO$^+$(2-1), and DCO$^+$(3-2) to further constrain the velocity structure.
For Oph~E as a whole, 
the three tracers indicate a mean velocity gradient $5.2 \pm 0.3$ km~s$^{-1}$~pc$^{-1}$ at
$P.A. = -108^\circ \pm 4^\circ$, in fairly good agreement with the gradient suggested by the 
N$_2$H$^+$(1-0) data (cf. Fig.~\ref{fig_oph_velcore}e).

In Oph~F, our N$_2$H$^+$(1-0) data indicate the presence of two distinct velocity components, 
overlapping near the position of F-MM2 in the plane of the sky (see Fig.~\ref{fig_oph_velcore}d).
Oph~F may thus actually 
consist of two independent dense cores, Oph~F1 and Oph~F2, 
with differing line-of-sight velocities ($\sim 4.2$ and $\sim 4.6$ km~s$^{-1}$).
Figure~\ref{fig_oph_velcore}d shows the direction of the mean
velocity gradient derived in each of these two cores (see Table~\ref{t:ophcef_gradv} for derived 
parameters).
The fact that the velocity gradient found in Oph~F1 
($\sim 3.8$ km~s$^{-1}$~pc$^{-1}$ at $P.A. \sim 17^\circ$) 
is nearly opposite to the gradient in Oph~F2 
($\sim 6.3$ km~s$^{-1}$~pc$^{-1}$ at $P.A. \sim -119^\circ$)
\footnote{The direction of the velocity gradient is however more uncertain in Oph~F2
since it was derived using only 5 independent points.}
supports the view that Oph~F is made up of two independent cores, 
partly overlapping in projection, and not a single core 
with a mean velocity gradient $\sim 1.8$ km~s$^{-1}$~pc$^{-1}$ 
at $P.A. \sim -50^\circ$ as suggested by \citet*{Loren90} 
based on lower-resolution DCO$^+$(2-1) observations.

For L1688 considered as a whole (cf. Fig.~\ref{fig_oph_velomap}), 
a global velocity gradient of $\sim 1.1$ km~s$^{-1}$~pc$^{-1}$ is seen from North-West 
(Oph~A) to South-East (Oph~F) (see also Loren 1989 in $^{13}$CO and Loren et al. 1990 in DCO$^+$).
The direction of this large-scale gradient is $P.A. \sim 120^\circ$ 
%Fig.~\ref{fig_oph_gradv} and   
(cf. note  \textit{a} of Table~\ref{t:oph_sigv}),
as determined by a least-squares fit to the observed  distribution of the line-of-sight 
systemic velocities of the condensations detected in N$_2$H\/$^+$(1-0). 
The observed velocity pattern is however clearly more complex than that expected from simple large-scale 
rotation \citep[see also ][]{Loren89b}. For instance, the direction of the mean velocity gradient
changes to $P.A. \sim 180^\circ$ if the condensations of Oph~B1 and Oph~B2 
are ignored. %(cf. Table~\ref{t:oph_sigv}).

\subsection{Relative motions of the protocluster condensations within the DCO\/$^+$ cores}
\label{sec_oph_vlsr}

\begin{figure*} [!t]
\vspace*{-0. cm}
\centerline{\resizebox{0.9\hsize}{!}{\includegraphics[angle=270]{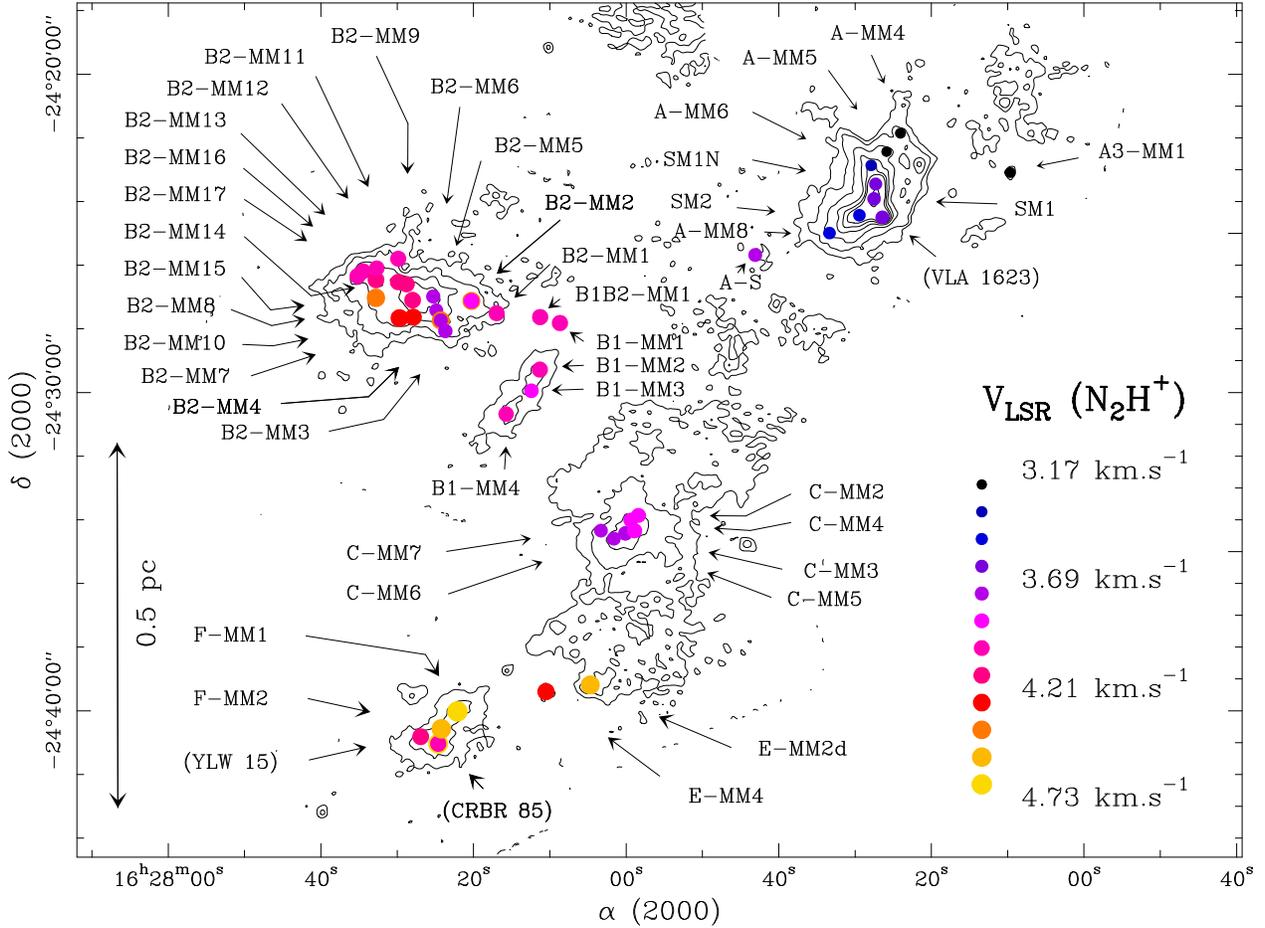}}}
\vspace*{-0.2 cm}
\caption{Line-of-sight systemic velocities of the  41 prestellar 
condensations  L1688 detected in N$_2$H\/$^+$(1-0), overlaid on the lowest contours 
of the 1.2~mm continuum mosaic of \citeauthor*{Motte98}. 
These velocities correspond to those listed in col.~[7] of  
Table~\ref{t:fithfs_results}.
Each condensation is represented by a filled circle whose size 
increases with $V_{LSR}$. The color coding varies from 
black to yellow 
with increasing Doppler shift. 
The systemic velocities of the protostars VLA~1623, YLW~15/IRS~43 and CRBR~85 
are also shown. A few condensations have two markers, 
reflecting the presence of two velocity components in their spectra.}
\label{fig_oph_velomap}
\end{figure*}

Our estimates of the centroid line-of-sight velocities toward the 
%$\rho$~Oph 
prestellar condensations  of L1688 (cf. col.~7 of Table~\ref{t:fithfs_results}) 
%(cf. col.~5 of Table~\ref{t:fithfs_results}) 
provide interesting constraints on the relative motions between condensations.
%, as well as on possible systematic large-scale motions within the protocluster. 
Figure~\ref{fig_oph_velomap} shows the distribution of systemic velocities as derived from
our N$_2$H$^+$(1-0) HFS fits, overlaid on the lowest contours of the 1.2~mm continuum map of 
\citeauthor*{Motte98}. 

\begin{table*} 
\centering 
\caption{Velocity dispersion of the L1688 protocluster condensations.} 
\label{t:oph_sigv} 
\vspace*{0.5ex} 
\begin{tabular}{lcccccccccc} 
\hline\hline 
Sample$^a$ & n$^b$ & D$^c$ & $<$\hspace*{-0.02em}$V_{lsr}^d$\hspace*{-0.15em}$>$ & $<$\hspace*{-0.02em}$V_{cent}^e$\hspace*{-0.15em}$>$ & $\sigma_{\mathrm{1D}}^f$ & $\sigma_{\mathrm{1D,c}}^g$ & $\sigma_{\mathrm{3D}}^h$ & $\sigma_{\mathrm{3D,c}}^i$ & D/$\sigma_{\mathrm{3D}}$ & D/$\sigma_{\mathrm{3D,c}}$ \\  
 & & {\tiny (pc)} & {\tiny (km~s$^{-1}$)} & {\tiny (km~s$^{-1}$)} & {\tiny (km~s$^{-1}$)} & {\tiny (km~s$^{-1}$)} & {\tiny (km~s$^{-1}$)} & {\tiny (km~s$^{-1}$)} & {\tiny ($10^6$ yr)} & {\tiny ($10^6$ yr)} \\  
(1) & (2) & (3) & (4) & (5) & (6) & (7) & (8) & (9) & (10) & (11) \\  
\hline 
Oph A  &  9  &        0.28  &        3.44  &        3.48  &        0.19(5)  &        0.20(5)  &        0.33(8)  &        0.34(9)  &         0.8  &         0.8  \\  
Oph B1/B2  &  24  &        0.33  &        4.05  &        3.96  &        0.20(3)  &        0.22(3)  &        0.35(5)  &        0.38(6)  &         1.0  &         0.9  \\  
Oph C,E,F  &  10  &        0.44  &        4.09  &        4.05  &        0.39(9)  &        0.39(9)  &        0.67(16)  &        0.68(16)  &         0.7  &         0.6  \\  
\hline 
L1688  &  47  &        1.10  &        3.95  &        3.90  &        0.36(4)  &        0.36(4)  &        0.62(7)  &        0.63(7)  &         1.8  &         1.7  \\  
L1688 $-\nabla{V}$ &  47  &        1.10  & - & - &        0.25(3)  & - &        0.43(4)  & - &         2.6  & -  \\  
\hline 
\end{tabular} 
\begin{list}{}{} 
%\item[]{Notes: the numbers in parentheses indicate the uncertainty in unit of the last digit.} 
\item[]{Notes: the numbers in parentheses indicate the uncertainty in units of the last digit.}
\item[$(a)$]{The first three samples contain the velocity components of Table~\ref{t:veldisp_results} only. 
The last two samples include the protostars in addition. 
For the\\ ''L1688 $-\nabla{V}$'' sample, the velocity dispersion is computed after removing the large scale velocity gradient 
measured with the method described in note $c$ of Table~\ref{t:ophcef_gradv}} using C-N as the (0,0) position: 
$V_0 =$       3.87 km~s$^{-1}$, $\|\vec{\nabla} V\| = $        1.1 km~s$^{-1}$pc$^{-1}$, $P.A. = $        117 $^\circ$. 
\item[$(b)$]{Number of velocity components used for the calculations.} 
\item[$(c)$]{Diameter of the region containing each sample.} 
\item[$(d)$]{Mean LSR velocity of the components, computed with the velocities listed in col.~[5] of Table~\ref{t:fithfs_results}.} 
\item[$(e)$]{Mean centroid velocity computed on the N$_2$H$^+$(101-012) spectra shown in Fig.~\ref{fig_oph_velcore}.} 
\item[$(f)$]{Standard deviation of the distribution of component LSR velocities around $<$\hspace*{-0.02em}$V_{lsr}$\hspace*{-0.15em}$>$ given in col.~[4]. 
The error bar was estimated as $\sigma_{\mathrm{1D}}/\sqrt{2(n-1)}$}, 
assuming that the sample is drawn from a larger population whose velocity distribution follows Gaussian statistics. 
\item[$(g)$]{Same as in col.~[6] but computed around $<$\hspace*{-0.02em}$V_{cent}$\hspace*{-0.15em}$>$ given in col.~[5].} 
\item[$(h)$]{3D velocity dispersion calculated from $\sigma_{\mathrm{1D}}$ assuming isotropic motions. The error bar was scaled from that estimated for $\sigma_{\mathrm{1D}}$.} 
\item[$(i)$]{Same as in col.~[8] but computed around $<$\hspace*{-0.02em}$V_{cent}$\hspace*{-0.15em}$>$ given in col.~[5].} 
\end{list} 
\end{table*}

Both the velocity differences between neighboring condensations and the overall 
velocity dispersion of the condensations within the 
L1688 protocluster are small. 
Table~\ref{t:oph_sigv}  gives the 
one-dimensional velocity dispersion,
$\sigma_{1D}$, derived from our N$_2$H\/$^+$(1-0) observations for various samples of condensations (cf. col.~[6] and col.~[7]).
This velocity dispersion was estimated in two ways, first around the mean LSR velocity of the condensations  
(estimate given in col.~[6]), and second around the mean centroid N$_2$H\/$^+$(101-012) velocity measured in the maps of 
Fig.~\ref{fig_oph_velcore} (estimate given in col.~[7]).
For the ensemble of 41 compact 
%[45 - C-N - C-W - E-MM1- B1B2-MM2 ] 
prestellar condensations and 3 protostars detected in N$_2$H$^+$, 
we estimate a global velocity dispersion $\sigma_{1D} \sim$ 0.36 km~s$^{-1}$ about 
the mean systemic velocity $<$\hspace*{-0.02em}$V_{lsr}$\hspace*{-0.15em}$> \sim$ 3.95 km~s$^{-1}$.
Assuming isotropic random motions, this corresponds to a three-dimensional velocity dispersion
$\sigma_{3D} = \sqrt{3}\ \sigma_{1D} \sim$ 0.62~km~s$^{-1}$ 
(cf. col.~[8] and [9] of Table~\ref{t:oph_sigv}). 
If the velocity distribution is Maxwellian, the mean condensation 
%velocity 
 speed relative to the center of mass of the system  
is $V_{mean} = \sqrt{8/\pi}\ \sigma_{1D} \sim$ 0.57~km~s$^{-1}$ 
and the mean relative  speed between condensations is 
$V_{rel} = \sqrt{2}\ V_{mean}  = \left(4/\sqrt{\pi}\right) \ \sigma_{1D} \sim$
0.81~km~s$^{-1}$.
Adopting a diameter $D \sim 1.1$~pc \citep[e.g.][]{Wilking83}  for the L1688 cloud,  
such a velocity dispersion implies a typical crossing time, 
$t_{cross} \equiv D /\sigma_{3D} \sim$ 1.8 $\times 10^6$~yr 
(cf. col.~[10] and col.~[11] of Table~\ref{t:oph_sigv}),
for the condensations within the L1688 protocluster. 
After subtracting the global, systematic velocity gradient of $\sim 1.1$ km~s$^{-1}$~pc$^{-1}$ 
seen across L1688 (cf. Sect.~\ref{sub_oph_gradv} above), 
the resulting 1D velocity dispersion of the condensations is even lower, 
$\sigma_{1D} \sim 0.25$ km~s$^{-1}$, suggesting that the crossing 
time associated with purely random condensation motions is larger, 
$t_{cross}  \sim 2.5 \times 10^6$~yr.
The crossing times estimated separately within the individual DCO$^+$ cores 
with a statistically significant number of condensations (i.e., Oph~A, B2, C) 
are only a factor of $\sim 2$ shorter, $t_{cross}  \sim$~0.6-1.0~$\times 10^6$~yr.

%\newpage
%%%%%%%%%%%%%%%%%%
%%
\section{Discussion}
\label{dis}

\subsection{Dynamical state and fate of the L1688 condensations}
\label{dis_vir}

The starless condensations identified by \citeauthor*{Motte98} 
in L1688 are highly 
centrally concentrated 
and feature large density contrasts over the local background medium, strongly suggesting 
they are self-gravitating. More specifically, the estimated mean densities of the condensations 
exceed the mean densities of the parent DCO$^+$ cores by a typical factor $\sim $~5--20, 
%by a typical factor $\sim $~10-20,  
%on average, 
while the mean column densities of the condensations exceed the background core 
column densities by a factor $\simgt $~2.
%by a factor $\sim $~2-4.
%$\sim 1.6$ to $\sim 4.7$. 
For comparison, a critical self-gravitating Bonnor-Ebert isothermal spheroid has a mean density 
contrast $\bar{\rho}_{\rm BE}/\rho_{\rm ext} \sim 2.4$
%of a factor $\sim 2.4$ 
\citep[e.g.][]{Lombardi01} and a mean column density contrast $\bar{\Sigma}_{\rm BE}/\Sigma_{\rm ext} \sim 1.5$
%of a factor $\sim 1.5$ 
over the external medium.

The line observations reported in this paper provide more direct evidence that most of the 
%$\rho$ Oph 
 L1688 condensations are gravitationally bound. 
%Importantly, 
Indeed, the narrow linewidths measured in N$_2$H$^+$(1-0) 
(see Sect.~\ref{obs_virial}) imply virial masses 
which generally agree within a factor of $\sim 2$ with the mass estimates derived 
by \citeauthor*{Motte98} from the 1.2~mm dust continuum. 
 This can be considered a good agreement since both $M_{vir}$ 
and $M_{1.2mm}$ are themselves uncertain by a factor of $\sim 2$. On the theoretical side,  
self-gravitating condensations are expected to have virial mass ratios  
$\alpha_{vir} \equiv M_{vir}/M_{1.2mm} \simlt 2$, 
while objects in gravitational virial equilibrium should have $\alpha_{vir} \sim 1$ 
\citep[e.g.][]{Bertoldi92}. 
 Here, among the 43 velocity components measured toward the 41 compact  
condensations positively detected in N$_2$H$^+$(1-0), 
15 have an estimated virial mass ratio 
$\alpha_{vir} < 1$ and another 23 have $\alpha_{vir}  \simlt 2.5$.
Altogether, 38 components have $\alpha_{vir} \simlt 2.5$ and 
only 5 have $\alpha_{vir} \simgt 3$. 
Among the latter 5 components, only 2 are 
associated with condensations more massive than 
$M_{1.2mm} \sim 0.35\, M_\odot$, 
of which one is associated with a condensation (B2-MM2) having another 
velocity component with $\alpha_{vir} < 2$.
Therefore, 37 of the 41 compact 1.2~mm condensations detected 
in N$_2$H$^+$(1-0) have at least one velocity component with 
$\alpha_{vir} \simlt 2.5$.

We conclude that a large majority ($\sim$ 75-90$\%$) 
%most ($\simgt 80\%$) 
of the  L1688 
condensations detected in N$_2$H\/$^+$(1-0), 
including $\sim$ 90$\%$ of those more massive 
than $\sim 0.35\, M_\odot $,  
are likely gravitationally bound\footnote{Note that the surface pressure term of 
the virial theorem, which we neglect here, tends to reduce the mass required for virial 
equilibrium, i.e., to increase the value of $\alpha_{vir}$ expected in virial equilibrium
-- see \citet{Bertoldi92}.}. 
Coupled with the detection of signatures of infall motions toward some of them 
(see Sect.~\ref{sub_oph_infall}), this supports the notion that 
$\sim 75\% $ of the starless condensations identified by \citeauthor*{Motte98} 
at 1.2~mm are {\it prestellar} in nature and on the verge of forming protostars 
 (cf. \citeauthor*{Andre00} 2000 and \citeauthor*{Ward06} 2007 for a definition of the prestellar stage).
%\citep[cf.][ and Ward-Thompson et al. 2007 for a definition of the prestellar stage]{Andre00}. 
A fraction ($\sim 30\% $) of the condensations with 
$M_{1.2mm} \simlt 0.35\, M_\odot$ may be only marginally bound.
The status of the condensations less massive than $\sim 0.1\, M_\odot $ is less clear 
since these are weaker and often undetected in N$_2$H$^+$. 
Some of them may possibly correspond to unbound transient objects generated by 
supersonic turbulence \citep[cf.][]{Klessen05}. % Ref Klessen et al. 2005
%(cf. Klessen et al. 2005). 
Interestingly, if a significant fraction of 
the condensations identified by \citeauthor*{Motte98} below 
$\sim 0.35\, M_\odot $ are indeed unbound and not truly prestellar in nature, 
then the apparent excess of starless condensations at the low-mass end of the mass spectrum 
compared to the IMF of field systems (cf. Fig.~1) may find a natural explanation 
in the context of a one-to-one mapping between prestellar condensations and stellar 
systems.

\subsection{Lifetime of the L1688 condensations}
\label{dis_lifetime}

The lifetime of the  L1688 prestellar condensations is rather uncertain but
several lines of reasoning suggest a typical value of $\sim 10^5$~yr. 
 Table~\ref{tab_timescales} provides estimates of some relevant evolutionary timescales. 
First, based on the measured mean densities $\bar{n}\h2 \equiv \bar{\rho}/\mu m_H \sim 10^5-10^7\, \rm{cm}^{-3}$, 
the free-fall dynamical timescales of the condensations, $t_{ff} \equiv \left(3\pi / 32G \bar{\rho} \right)^{1/2} $,  
range from $t_{ff} \sim 10^4$~yr to  $t_{ff} \sim 10^5$~yr. 
While some of the condensations show evidence of infall motions, most of them are 
unlikely to be in free-fall collapse. We may conservatively estimate their evolutionary
timescale to be $\sim 3\, t_{ff}$, which is appropriate for magnetically supercritical 
dense cores \citep[e.g.][]{Ciolek94} and typical of isolated prestellar cores detected 
in the submillimeter continuum \citep[][]{Kirk05}. 
This leads to timescale values ranging from  $\sim 2 \times 10^4$~yr to $\sim 3.5 \times 10^5$~yr 
(cf. col.~[6] of Table~\ref{tab_timescales}).
Second, assuming that the  L1688 condensations evolve into stellar systems at a constant 
rate, we may derive a rough statistical estimate of their lifetime (col.~[7] of Table~\ref{tab_timescales})
by comparing the number of observed prestellar condensations (col.~[3] of Table~\ref{tab_timescales})
to the number of pre-main sequence (PMS) systems found in the same region (col.~[2] of Table~\ref{tab_timescales}). 
The ISOCAM mid-IR survey of \citet{Bontemps01} revealed a total of 109 Class~II PMS  
objects in L1688 with ages $\sim 0.4-1$~Myr. If all of the 45 starless condensations identified 
by \citeauthor*{Motte98} above $0.1\, M_\odot $ are truly prestellar in nature, this points 
to a prestellar lifetime $\sim 2-4 \times 10^5$~yr. 
A third timescale estimate may be obtained by dividing the typical condensation outer radius 
$r_{cond} \sim 4000$~AU  by a typical infall speed $V_{inf} \sim 0.1-0.3$~km/s \citep[cf.][]{Belloche01}. 
This approach yields a condensation lifetime $\sim 0.6-2 \times 10^5$~yr.
We conclude that the 1.2mm continuum condensations  of L1688 are likely characterized 
by a range of lifetimes between  $\sim 2 \times 10^4$~yr and $\sim 5 \times 10^5$~yr 
(see col.~[6] and col.~[7] of Table~\ref{tab_timescales}).

\begin{figure} [!ht]
\centerline{\resizebox{1.0\hsize}{!}{\includegraphics[angle=270]{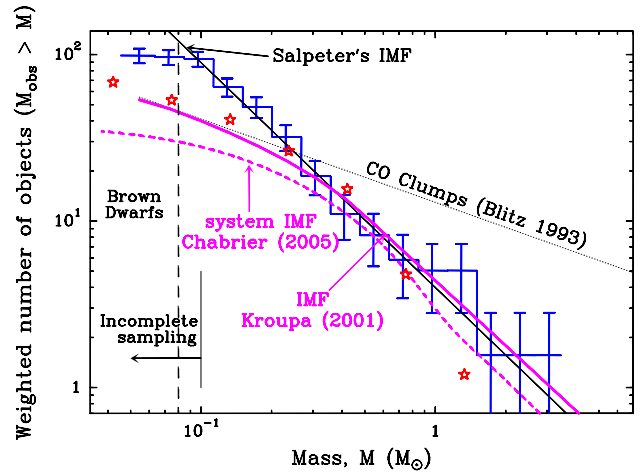}}}
%\vspace*{-0.15 cm}
\caption[Weighted mass spectrum of $\rho$~Oph starless condensations]
{Weighted cumulative mass spectrum of the 57 starless condensations identified by \citeauthor*{Motte98}
(histogram with error bars), compared to the same mass distributions as shown in Fig.~\ref{fig_preimf}, 
as well as the Salpeter power-law IMF (solid line).
Here, each condensation was assigned a weight inversely proportional to its estimated 
free-fall dynamical timescale (see text). 
The flattening apparent below $\sim 0.4\, M_\odot $  in the unweighted CMD 
(Fig.~\ref{fig_preimf}) is not seen in the weighted CMD shown here, which is essentially consistent 
with a single, Salpeter-like power law.}
\label{fig_wcmd}
\end{figure} 

As pointed out by \citet{Clark07},  
%Clark, Klessen, \& Bonnell (2007), 
if the lifetime of the condensations depends 
on their mass, then the observed mass spectrum is not necessarily representative of the intrinsic 
condensation mass distribution (CMD) \citep[see also][]{Elmegreen00}. 
This is due to the fact that an observer 
is more likely to detect long-lived condensations than short-lived condensations. 
Here, however, the mean densities of the  L1688 condensations are essentially uncorrelated with
their masses, so that there is no systematic dependence of the dynamical timescale on the mass.
To quantify the importance of the potential timescale bias, we plot, in Fig.~\ref{fig_wcmd}, 
a weighted version of the  central Ophiuchus CMD in which each condensation was assigned 
 a weight equal to $\frac{<t_{ff}>}{t_{ff}} = \frac{\bar{\rho}^{1/2}}{<\bar{\rho}^{1/2}>}$ (instead of 1 as used for 
Fig.~\ref{fig_preimf}),  where $<t_{ff}>$ is
%$<\bar{\rho}^{1/2}>$ are 
the average free-fall time of the condensations. 
%a weight proportional to the square root of its estimated mean density.  
Such a weighting should allow us to recover the intrinsic shape of the CMD assuming that the lifetime of 
each condensation is proportional to its free-fall time.
As  can be seen in Fig.~\ref{fig_wcmd}, this weighting does not change the high-mass end of the CMD 
and only affects the low-mass end. We conclude that the steep, Salpeter-like slope of the CMD 
at the high-mass end is robust, but that the presence of a break at $\sim 0.4\, M_\odot $ is 
less robust.

\begin{table*} 
\vspace*{2.ex}
 \caption[Timescales of the $\rho$~Oph prestellar condensations]{Evolutionary timescales 
for various systems of protocluster condensations.
}
 \label{tab_timescales}
 %\vspace*{2.ex}
 \centering
 \footnotesize
 \begin{tabular}{lcccccccc}
 \hline
 \hline
\multicolumn{1}{c}{Sub-cluster} & \multicolumn{1}{c}{Nb of} & \multicolumn{1}{c}{Nb of starless } & \multicolumn{1}{c}{$1+\Theta$} & \multicolumn{1}{c}{$\frac{R^2}{N_{cond}r_{cond}^2}$} & \multicolumn{1}{c}{$3\, t_{ff}$} & \multicolumn{1}{c}{Statistical} & \multicolumn{1}{c}{$ t_{cross}$} & \multicolumn{1}{c}{$ t_{coll}$}   \\
\multicolumn{1}{c}{ } & \multicolumn{1}{c}{Class II} & \multicolumn{1}{c}{condensations} & \multicolumn{1}{c}{ } & \multicolumn{1}{c}{  } & \multicolumn{1}{c}{ } & \multicolumn{1}{c}{lifetime}  & \multicolumn{1}{c}{ }                                   & \multicolumn{1}{c}{ }   \\
\multicolumn{1}{c}{ } & \multicolumn{1}{c}{YSOs} & \multicolumn{1}{c}{$> 0.1\, M_\odot $ } & \multicolumn{1}{c}{} & \multicolumn{1}{c}{} & \multicolumn{1}{c}{($10^5$~yr)} &  \multicolumn{1}{c}{($10^5$~yr)} & \multicolumn{1}{c}{($10^5$~yr)} & \multicolumn{1}{c}{($10^5$~yr)}   \\*[0.4ex]
 \hline
 L1688              &   109 &  45 & 2	  & 35 & 0.2-3.5  &   2-4        &  18	& 160	   \\
 Oph~A              &    41 &   9 & 8	  & 10 & 0.5-2    & 0.9-2        &  8	&  5.5	    \\
 Oph~B1/B2          &    31 &  22 & 3.5   & 15 & 0.2-2.5  &   3-7        &   10   & 22	    \\
 Oph~C/E/F          &    35 &  12 & 3	  & 15 & 0.3-3.5  &   2-4        &    7    & 19	    \\
 \hline
 \end{tabular}
 \normalsize
\end{table*}

\subsection{Likelihood of interactions between condensations}
\label{dis_interactions}

Using the velocity dispersions derived for the condensations in Sect.~\ref{sec_oph_vlsr} and 
the lifetime estimates of Sect.~\ref{dis_lifetime} above, we are now in a position to assess 
whether dynamical interactions between condensations are a frequent or rare phenomenon 
in the  L1688 protocluster.

Assuming an isotropic, Maxwellian velocity distribution, 
the time, $t_{coll}$, required for a condensation to interact or collide with another
condensation can be estimated from the following formula for the collision rate 
$1/t_{coll}$ :

\begin{equation}
1/t_{coll} = 4\, \sqrt{\pi} \times n_{cond}\, \sigma_{1D}\, r_{cond}^2 \times 
%\left(1+\frac{GM_{cond}}{2 \sigma_{1D}^2 r_{cond}} \right), $$
\left(1+\Theta \right), 
\end{equation}

\noindent
where $n_{cond}$ is the number density of condensations, 
$r_{cond}$ is the condensation outer radius (which we take to be 
twice the measured HWHM radius at 1.2mm - cf. Sect.~\ref {obs_virial}), 
$\pi  r_{cond}^2 $ is the collisional cross section 
of the condensation, and $1+\Theta \equiv 1+GM_{cond}/(\sigma_{1D}^2 r_{cond})$ 
is the gravitational focusing factor \citep[e.g.][]{Binney87}.
Further assuming that the condensations are homogeneously distributed in a spherical system 
of radius $R$, so that $ n_{cond} = N_{cond}/(\frac{4}{3}\pi R^3) $, the ratio 
of $t_{coll}$ to $t_{cross}$ takes on the simple form:

\begin{equation}
\frac{t_{coll}}{t_{cross}} = \frac{1}{2} \sqrt{ \frac{\pi}{3} } \times 
\frac{R^2}{N_{cond}\, r_{cond}^2} \times \frac{1}{1+\Theta}.
\end{equation}

Using the above equation for the entire L1688 cluster-forming clump ($R \sim 0.55$~pc, $N_{cond} = 57$, $\sigma_{1D} = 0.36$~km~s$^{-1}$) 
and adopting typical condensation properties ($r_{cond} \sim 2500$~AU and $M_{cond} \sim 0.4\, M_\odot $), we find 
$t_{coll}^{L1688}/t_{cross}^{L1688} \sim 9$ and thus $t_{coll}^{L1688} \sim 16 \times 10^6$~yr. 
The latter is two orders 
of magnitude larger than the estimated lifetime of the condensations (Sect.~\ref{dis_lifetime}) 
and about one order of magnitude longer than the age of the  L1688 PMS cluster 
\citep{Bontemps01}. 
 For simplicity, our analysis ignores the dynamical impact of the ambient protocluster environment.
In such a dense environment, gas drag is likely to have an effect on the motions of individual condensations, 
which will make the associated velocity field depart from a true Maxwellian distribution. A detailed assessment 
of the effects of gas drag is beyond the scope of this paper, but qualitatively at least we expect the environment 
to make the interaction timescales even longer than the above estimates.
We conclude that 
the interaction process is much too slow on the scale of the entire L1688 
system to play any significant role in the evolution of the protocluster condensations.

Dynamical interactions are however more likely to occur on smaller scales than the whole L1688 cluster-forming clump.
In their mid-IR census of the PMS cluster, \citet{Bontemps01} identified three main sub-clusters associated with 
the DCO$^+$ cores Oph~A, Oph~B1/B2, and Oph~C/E/F, respectively. The fact that sub-clustering is observed in the spatial 
distribution of PMS objects further supports our conclusion that significant interactions between individual objects 
have not yet occurred on the scale of the entire protocluster. 
Using Eq.~(2) for the condensations associated with each of the three sub-cluster 
systems ($R \sim 0.15$~pc, $N_{cond} = 10-20$), 
we find shorter interaction timescales than for L1688 as a whole, 
$t_{coll}^{sub} \sim 6-22 \times 10^5$~yr  (see col.~[9] of Table~\ref{tab_timescales}). 
The derived collision timescales nevertheless remain significantly longer than the 
typical condensation lifetime $\sim 10^5$~yr, 
indicating that dynamical interactions between condensations cannot be a dominant process even on the scale of the 
sub-clusters and DCO$^+$ cloud cores. 
The collision timescales are even longer than the combined lifetime of the prestellar and protostellar (Class~0/Class~I) 
phases, which we estimate to be at most $\sim 5 \times 10^5$~yr \citep[e.g.][]{Greene94}.
In general, therefore, the prestellar condensations do not have 
time to orbit through their parent  DCO$^+$ core and collide with one another before evolving into Class~II PMS objects. 
We note that $t_{coll}$ is only slightly larger than $ 5 \times 10^5$~yr  
in the case of the Oph~A 
cluster-forming core, suggesting that a few interactions may occur during the whole evolution of this sub-cluster. 
Interestingly, based on a recent detailed study of the NGC~2264-C cluster-forming clump in the Mon~OB1 complex, 
\citet{Peretto06} found direct evidence of a strong dynamical interaction in the dense inner region of that 
protocluster. However, the interaction observed at the center of NGC~2264-C is purely gravitational in origin and 
results from large-scale, coherent collapse motions as opposed to random turbulent motions \citep{Peretto07}.

Our present results in  L1688 seem inconsistent with models which resort to 
strong dynamical interactions 
%and competitive accretion 
to build up a mass spectrum comparable to the IMF \citep[e.g.][ -- see also Sect.~\ref{dis_accretion} below]{Price95}.

\begin{table*}
\vspace*{2.ex}
 \caption[Rates of competitive mass accretion]{Rates of competitive mass accretion 
for a range of objects in the  L1688 protocluster.}
 \label{accretion}
% \vspace*{2.ex}
 \centering
 \footnotesize
 \begin{tabular}{lcccc}
 \hline
 \hline
\multicolumn{1}{c}{Object} & \multicolumn{1}{c}{$n_{\rm H2}^{back}$ } & \multicolumn{1}{c}{$v_{rel}$ }           & \multicolumn{1}{c}{$R_{acc}$} & \multicolumn{1}{c}{$\dot{M}_{acc} $}   \\
\multicolumn{1}{c}{}            & \multicolumn{1}{c}{(${\rm cm}^{-3}$) }        & \multicolumn{1}{c}{(km~s$^{-1}$) }  & \multicolumn{1}{c}{(AU)}           & \multicolumn{1}{c}{($M_\odot \, {\rm yr}^{-1} $)}    \\*[0.4ex]
 \hline 
 Class I in L1688                  &   $2 \times 10^4$                                             &   0.4                                                        & 3500                                              & $4 \times 10^{-7}$	   \\
 Class I in DCO$^+$ core  &   $5 \times 10^4$                                             &   0.3                                                        & 3000                                               &  $6 \times 10^{-7}$	   \\
 Class 0 in inner Oph A core  &   $4 \times 10^5$                                        &   0.3                                                        & 2500                                               &  $3 \times 10^{-6}$	   \\
 Class 0 from multiple system          &   $10^6$                                        &   0.2                                                        & 2000                                               &  $3.5 \times 10^{-6}$	   \\
 ~inside collapsing condensation  &                                                    &                                                                &                                                           &    	   \\
 Condensation in inner Oph A core  &   $4 \times 10^5$                             &   0.3                                                        & 2500                                               &  $3 \times 10^{-6}$           \\
 \hline
 \end{tabular}
 \normalsize
\end{table*}

\subsection{Comparison with the competitive accretion picture}
\label{dis_accretion}

In  a scenario of clustered star formation proposed by 
Bonnell, Bate and collaborators \citep[e.g.][]{Bonnell01a,Bonnell01b}, 
the IMF is primarily determined by competitive accretion and dynamical 
interactions/ejections during the protostellar phase, corresponding observationally to  
Class~0 and Class~I objects. In this picture, turbulence generates density fluctuations 
within molecular clouds, some of which are gravitationally unstable and collapse to 
interacting protostars or protostellar seeds. 
The initial envelope/core masses of these protostars are essentially uncorrelated with the 
corresponding final stellar masses, especially at the high-mass end of the IMF 
\citep[cf.][]{Bonnell04}.
The protostars acquire most of their mass by moving around in the 
%(gas-dominated) 
gravitational potential well of the parent cluster-forming clump and accreting background gas  
that initially did not belong to the corresponding protostellar envelopes/cores.
This process of competitive accretion of background gas is highly non-uniform and depends 
primarily on the initial stellar position $R_\star$ within the protocluster.
The few protostars initially located near the center of the 
cluster potential accrete rapidly from the start and 
become massive stars, while protostars in the low-density 
outer regions accrete much more slowly and become low-mass stars 
\citep[see][ for a quantitative toy model consistent with the observed IMF]{Bonnell01b}. 

In the context of this scenario, it is difficult to explain why  the Ophiuchus  
prestellar condensations have a mass distribution resembling the IMF. 
Our present results on the kinematics 
of the protocluster condensations allow us to further discuss the possible relevance of 
the competitive accretion picture in the case of the  L1688 embedded cluster.
The rate at which a protostar accretes mass competitively as it travels through 
the background protocluster gas is $\dot{M}_{acc} \approx \pi\, \rho_{back} \, v_{rel}\, R^2_{acc} $ 
\citep[cf.][]{Bonnell01a}, which gives:
 
\begin{equation}
\dot{M}_{acc}  \sim 1.2 \times 10^{-6}\, M_\odot {\rm yr}^{-1}  
 \left(\frac{n_{\rm H2}^{back} }{ 10^5\, {\rm cm}^{-3}} \right) \left(\frac{v_{rel}}{0.3\, {\rm km/s}} 
 \right) \left(\frac{R_{acc}}{3000\, {\rm AU}}\right)^2,
\end{equation}

\noindent
where $\rho_{back}$ is the background gas density, $v_{rel}$ is the relative
velocity between the protostar and the local ambient gas, and $R_{acc}$ is the accretion radius. 
\citet{Bonnell01a} have shown that a good analytic approximation for $R_{acc}$ 
is provided by the smaller of the Bondi-Hoyle radius, $R_{BH} = 2GM_\star / \left(v_{rel}^2+c_s^2\right)$, 
and the local tidal-lobe radius, 
$R_{tidal} \approx 0.5  \left(\frac{M_\star}{M_{enc}}\right)^{1/3} R_\star$, 
where $c_s$ is the gas sound speed and $M_{enc}  \left(R_\star \right)$ is the mass enclosed within the protocluster at the
protostar's position $R_\star$. 
The Bondi-Hoyle radius, $R_{BH}$, is the radius where the gravitational potential due to the protostar
exceeds the kinetic energy of the gas. Here, we estimate the relative gas--protostar velocity to be 
the derived mean condensation  speed $V_{mean} = \sqrt{8/\pi}\ \sigma_{1D} \sim 0.3-0.4$~km~s$^{-1} \simlt 2\, c_s$, 
assuming a Maxwellian velocity distribution. Therefore, the typical value of the Bondi-Hoyle radius is:
$R_{BH} \sim 9900\, {\rm AU}\, \left(M_\star /0.5\, M_\odot \right) \left(v_{rel} /0.3\, {\rm km/s} \right)^{-2}$.\\
The tidal radius,  $R_{tidal}$, expresses the fact that the tidal forces exerted by the 
gravitational potential of the ambient protocluster limit the zone of influence 
of a given protostar. 
In gas-dominated protoclusters such as L1688, one usually has $R_{tidal} \simlt R_{BH} $ and 
thus $R_{acc} \approx R_{tidal}$;  conversely, in stellar-dominated clusters 
$R_{BH}  \simlt R_{tidal}$ and thus $R_{acc} \approx R_{BH}$ \citep[][]{Bonnell01a}.

To estimate the typical value of $R_{tidal}$ in the  central Ophiuchus case, we assume that the overall gas distribution in 
the centrally-condensed L1688 protocluster and individual subclusters (e.g. Oph~A)  
follows a $\rho \propto r^{-2}$ density profile on average. 
This assumption is consistent with  the density gradient derived by \citeauthor*{Motte98}
for the outer parts of the DCO$^+$ cores based on the circularly-averaged profiles observed 
at 1.2mm. 
Under this assumption,  $M_{enc}$ has a very simple 
expression, $M_{enc}  \left(R_\star \right) =  M_{clus} \times \left(R_\star/R_{clus}\right)$, 
and thus $R_{tidal} \approx 0.5 \left(\frac{M_\star}{M_{clus}}\right)^{1/3} R_{clus}^{1/3}\,  R_\star^{2/3}$, 
where $M_{clus}$ and $R_{clus}$ are the total  (gas $+$ stars) mass and outer radius of the 
protocluster, respectively. 
For the entire L1688 protocluster, we have  $M_{clus} \sim 650\, M_\odot$, $R_{clus} \sim 0.55$~pc, 
and thus $R_{tidal}^{L1688} \sim 3500\, {\rm AU}\, \left(M_\star /0.5\, M_\odot \right)^{1/3} 
\left(R_\star /0.3\, {\rm pc} \right)^{2/3}$.
For the Oph~A subcluster, we adopt $M_{clus} \sim 30\, M_\odot$, $R_{clus} \sim 0.14$~pc, 
and thus $R_{tidal}^{OphA} \sim 3000\, {\rm AU}\, \left(M_\star /0.5\, M_\odot \right)^{1/3} 
\left(R_\star /0.1\, {\rm pc} \right)^{2/3}$.
As expected, we see that  $R_{tidal} < R_{BH} $ and thus $R_{acc} \approx R_{tidal} \sim 3000\, {\rm AU} $. 

Based on these simple estimates, we find that the typical tidal-lobe radius of the 
%$\rho$~Oph 
 L1688 condensations/protostars is comparable to the actual radius of the condensations 
and protostellar envelopes as measured by \citeauthor*{Motte98} at 1.2~mm. 
This suggests that the overall gravitational potential of the  L1688 protocluster 
does play an important role in limiting 
the condensation/envelope masses, in qualitative agreement with the competitive accretion 
picture \citep[cf.][]{Bonnell01a,Bonnell04}. Note that the small value found here for 
the mean condensation velocity relative to the local gas ($V_{mean}  \sim 0.3$~km~s$^{-1}$) 
is also consistent with the predictions of the competitive accretion model during the gas-dominated
phase of protocluster evolution \citep[][]{Bonnell01a}. 
However, when we estimate the mass accretion rate resulting from competitive accretion 
in various typical situations, we find relatively low values (see Table~\ref{accretion}), 
%$\simlt 3 \times 10^{-6}\, M_\odot \, {\rm yr}^{-1}$ 
which are smaller than the infall rate expected from the gravitational collapse 
of individual condensations, i.e., $\dot{M}_{inf} \sim 1-10\, c_s^3/G $ 
(see \citeauthor{Shu77} 1977 and \citeauthor{FC93} 1993 for model predictions, 
and \citeauthor{Belloche06} 2006 for an observed example). 
For a Class~0 protostar such as VLA~1623 embedded in the inner part of the Oph~A DCO$^+$ core, we estimate 
the background gas density to be relatively high, $n_{\rm H2}^{back} \sim 4 \times 10^5\,  {\rm cm}^{-3} $, 
and the tidal radius to be at most
$R_{tidal}^{VLA1623} \sim 2500\, {\rm AU}$, giving $ \dot{M}_{acc} \sim 3 \times 10^{-6}\, M_\odot \, {\rm yr}^{-1}$. 
For comparison, the 
infall rate due to gravitational collapse is significantly higher at this early stage 
$\dot{M}_{inf} \simgt 10\, c_s^3/G \simgt 10^{-5}-10^{-4}\, M_\odot \, {\rm yr}^{-1}$ 
\citep[][]{Bontemps96,Jayawardhana01,Andre01}.
For a more evolved Class~I protostar such as YLW~15/IRS~43 embedded in the Oph~F DCO$^+$ core, 
the background gas density 
is lower, $n_{\rm H2}^{back} \sim 5 \times 10^4\,  {\rm cm}^{-3} $, and the tidal radius is at most
$R_{tidal}^{IRS43} \sim 3000\, {\rm AU}$, giving 
$ \dot{M}_{acc} \sim 6 \times 10^{-7}\, M_\odot \, {\rm yr}^{-1}$. For comparison, 
the gravitational infall rate is $\dot{M}_{inf} \sim c_s^3/G \sim 2 \times 10^{-6}\, M_\odot \, {\rm yr}^{-1}$ 
at this late protostellar stage 
\citep[e.g.][]{Adams87,Bontemps96}. For both Class~0 and Class~I objects we thus find 
$ \dot{M}_{acc} < \dot{M}_{inf}/3$, implying 
that local gravitational collapse dominates over competitive accretion
\citep[see also][]{Krumholz05}.
We conclude that competitive accretion at the protostellar stage 
is unlikely to be the main mechanism responsible 
for determining the final masses of stellar systems in the  central Ophiuchus star-forming cloud. 
If each prestellar condensation fragments into a small-N system 
during protostellar collapse \citep[e.g.][]{Goodwin06}, %Ref Goodwin et al. 2006 ?
%(e.g. Goodwin et al. 2006), 
then competitive accretion may possibly play 
a more important role in defining the final masses of the individual components.

We also believe that competitive, Bondi-like accretion is more likely to operate {\it at the prestellar stage} and may 
possibly govern the growth of starless condensations within a cluster-forming
cloud \citep[cf.][]{Bonnell01b,Myers00a,Basu04}. 
%\citep[cf.][]{Bonnell01b,Myers00a}. 
The growth rate of a typical $\sim 0.5\, M_\odot $ condensation embedded in the inner part of a DCO$^+$ core such as Oph~A
should be similar to the competitive accretion rate estimated above for VLA~1623, i.e., 
$ \dot{M}_{acc} \sim 3 \times 10^{-6}\, M_\odot \, {\rm yr}^{-1}$. This is sufficient to approximately double the mass 
of the condensation in $\sim 2 \times 10^5$~yr, a timescale comparable to the condensation lifetime. 
However, once fast, nearly free-fall collapse sets in, we expect the growth rate to be quickly overwhelmed by 
the infall rate, so that a collapsing prestellar condensation should not have time to grow significantly in mass before 
evolving into a PMS system.

Another feature of the dynamics of a gas-dominated protocluster 
is that the entire system is expected to undergo global collapse/contraction, 
resulting in a centrally-condensed overall structure much like a self-gravitating
isothermal sphere \citep[cf.][]{Adams00,Bonnell01b}. % Ref Adams 2000 ?
%(cf. Adams 2000 and Bonnell et al. 2001a).
In this view, 
both the gas accretion and the protocluster evolution occur on the global dynamical  
timescale \citep[see also][]{Klessen00}.\\ 
%
%\subsection{Dynamical state of the protocluster}
%\label{dis_dynamics}
%
Observationally, the velocity dispersion, $\sigma_{1D}$, estimated in Sect.~\ref{sec_oph_vlsr} 
above for the  L1688 condensations implies a binding virial mass 
$M_{vir} \approx 3 \times R\sigma_{1D}^2/G \sim 50\, M_\odot$, which is much less than 
the total gas mass $\sim 550\, M_\odot $ of the associated C$^{18}$O cloud 
\citep[cf.][]{Wilking83} and even less than the total stellar mass of 
the present infrared embedded cluster 
\citep[$M_{\mbox{stars}} \sim 100\, M_\odot$,][]{Bontemps01}. 
In other words, the observed condensation-to-condensation velocity dispersion is 
a factor of $\sim 3$ smaller than that expected in virial equilibrium \citep[see also Table~6 of][]{Peretto06},
Since the magnetic field does not seem to be strong enough to support the 
cloud  \citep[e.g.][]{Troland96}, this comparison suggests that {\it the  L1688 system is indeed
gravitationally unstable and possibly in an early state of large-scale, magnetically supercritical 
%global
contraction}. 
Interestingly, the CO and $^{13}$CO lines 
observed toward the  L1688 cloud exhibit the classical spectroscopic
signature of contraction motions (cf. Sect.~\ref{sub_oph_infall} above) over most of the 
protocluster extent \citep*[see][]{Encrenaz75}. 
We also note that L1688 is not the only protocluster for which both subvirial relative speeds and evidence 
of large-scale contraction have been found. Two other examples are NGC~2264 
\citep[][]{Peretto06,Peretto07} and NGC~1333 \citep[][]{Walsh06,Walsh07}. 
This suggests that protoclusters often start their evolution from ``cold'', out-of-equilibrium 
initial conditions \citep[cf.][]{Adams06}, possibly as a result of the influence of external triggers \citep[cf.][]{Nutter06}.

\section{Summary and conclusions}
\label{summary}

We carried out a detailed observational study of the kinematics of the  L1688 protocluster 
condensations in the central Ophiuchus molecular cloud 
using N$_2$H\/$^+$(1-0) as a tracer of dense gas.  Additional observations 
were also taken in molecular lines such as H$^{13}$CO$^+$(1-0),  DCO$^+$(2-1), HCO$^+$(3-2), 
CS(2-1), and C$^{34}$S(2-1).
Our main results and conclusions are as follows:

\begin{enumerate}

\item The N$_2$H\/$^+$(1-0) line was positively detected toward  41 of the 57 compact starless 
condensations identified by \citeauthor*{Motte98} at 1.2~mm, as well as 3 Class~0/Class~I protostars 
(VLA~1623, CRBR~85, YLW~15/IRS~43). The same objects were also detected in H$^{13}$CO$^+$(1-0)
and/or DCO$^+$(2-1) when observed in these transitions. In addition, the Class~I sources LFAM~26 
and GY210 were detected in H$^{13}$CO$^+$(1-0).

\item  For 29 of the 44  
%For 32 of the 48 
condensations/protostars detected in N$_2$H\/$^+$(1-0) we are confident that the line emission 
is at least partly associated with the compact 1.2~mm continuum object  as opposed to the more extended 
parent cloud/DCO$^+$ core, since the sources remained positively detected after subtraction of the local 
background line emission. Furthermore, at least 17 starless condensations and 3 protostars were found 
to have well-defined line counterparts in position-velocity space and could be identified in our 
background-subtracted N$_2$H$^+$(101-012), H$^{13}$CO$^+$(1-0), or DCO$^+$(2-1) data cubes 
using the {\it Gaussclumps} clump-finding algorithm of \citet{Stutzki90}.

\item The measured N$_2$H\/$^+$(1-0) linewidths are narrow, indicative of 
small nonthermal velocity dispersions. The condensations of Oph~A, Oph~B1, Oph~C, Oph~E, and Oph~F 
are characterized by {\it subsonic} levels of internal turbulence ($\sigma_{NT}/\sigma_{T} < 1$), while  
the condensations of Oph~B2 have at most ``transonic'' internal turbulence ($\sigma_{NT}/\sigma_{T}  < 2$). 

\item We detected the classical spectroscopic signature of infall motions toward six 1.2~mm continuum 
condensations (A-SM2, B2-MM16, C-MM5, C-MM6, E-MM2d, and E-MM4). For these objects, 
the optically thick CS(2--1), CS(3--2), H$_2$CO($2_{12} - 1_{11}$), and/or HCO$^+$(3--2) lines  are double-peaked 
with a stronger blue peak, while the low-optical-depth N$_2$H$^+$(101--012) line peaks 
in the dip of the optically thick line. The same signature was tentatively observed in 10 other condensations. 

\item The virial masses derived from our N$_2$H$^+$(1-0) detections of the condensations 
generally agree within a factor of $\sim 2$ with the masses estimated from the 1.2~mm dust continuum. 
On this basis, most ($\sim $~75-90$\%$) of the L1688 condensations detected in N$_2$H\/$^+$(1-0), 
including essentially all ($\sim  90\% $) of those more massive than $\sim 0.35\, M_\odot $,  
appear to be gravitationally bound and prestellar in nature.
The status of the starless 1.2~mm condensations less massive than $\sim 0.1\, M_\odot $ is less clear 
since these were often undetected in N$_2$H$^+$. 

\item We estimate that the prestellar condensations of L1688 are characterized by a range of lifetimes 
between $\sim 2 \times 10^4$~yr and $\sim 5 \times 10^5$~yr. There is however no systematic dependence 
of lifetime on mass, so that the observed mass spectrum is not severely affected by the timescale bias pointed out 
by \citet{Clark07}. In particular, the steep, Salpeter-like slope of the condensation mass distribution at 
the high-mass end appears to be robust.

\item Based on the observed distribution of N$_2$H$^+$(1-0) line centroid velocities, a global 
velocity dispersion $\sigma_{1D} <  0.4$ km~s$^{-1}$ was estimated for the condensations of L1688.
This condensation-to-condensation velocity dispersion is subvirial. 
It implies a crossing time $t_{cross}  \sim 1$~Myr for the condensations 
within the parent protocluster and a typical collision time $t_{coll}  \sim 1-10$~Myr between condensations.
%Such a small velocity dispersion implies a crossing time $t_{cross}  \sim 1$~Myr for the condensations 
%within their parent protocluster and a typical collision time $t_{coll}  \sim 1-10$~Myr between condensations.
Since these timescales are longer than the estimated condensation lifetime $\sim$~0.2-5~$\times 10^5$~yr, we 
conclude that, in general, the L1688 prestellar condensations 
do not have time to orbit through their parent  
DCO$^+$ core and interact with one another before evolving into PMS objects. 

\item Using an estimated mean relative velocity of $\sim 0.3-0.4$~km~s$^{-1}$ between protostellar envelopes  
and the background ambient gas, the mass accretion rate expected from competitive accretion was found to be 
significantly smaller than that resulting from gravitational collapse at the Class~0 stage. On the other hand, 
the typical tidal-lobe radius of the L1688 condensations/protostars was estimated to be comparable to the 
actual radius of the condensations and protostellar envelopes as measured in the dust continuum at 1.2~mm. 
This suggests that the overall gravitational potential of the protocluster does play an important role 
in limiting the condensation/envelope masses, but that competitive accretion at the protostellar stage 
is not the dominant mechanism responsible for determining the final masses of stellar systems 
in the  central Ophiuchus star-forming cloud. 

\item We find that competitive, Bondi-like accretion is more likely to operate at the prestellar stage and 
propose that it  may partly govern the growth of starless, self-gravitating condensations initially produced 
by gravoturbulent fragmentation toward an IMF, Salpeter-like mass spectrum.

\end{enumerate}

\begin{acknowledgements} 
We would like to thank Bruce Elmegreen and Ian Bonnell for enlightening discussions on  
protocluster dynamics.  We are grateful to an anonymous referee for constructive comments 
that helped us improve the clarity of the paper.
\end{acknowledgements}

\end{document}